\documentclass[fleqn,usenatbib]{mnras}

\usepackage{newtxtext,newtxmath}

\usepackage[T1]{fontenc}

\DeclareRobustCommand{\VAN}[3]{#2}
\let\VANthebibliography\thebibliography
\def\thebibliography{\DeclareRobustCommand{\VAN}[3]{##3}\VANthebibliography}

\usepackage{graphicx}	
\usepackage{amsmath}	
\usepackage{booktabs}
\usepackage{pdflscape}

\usepackage{longtable}

\title[CSPNe extinctions]{Central-star extinctions towards planetary nebulae}

\author[A. Csukai et al.]{
Alexander Csukai$^{1,2}$,
Albert A. Zijlstra$^{1,2}$,
Iain McDonald$^{1}$
and Orsola De Marco$^{2,3}$
\thanks{Contact e-mail: \href{alexander.csukai@manchester.ac.uk}{alexander.csukai@manchester.ac.uk}}
\\
$^{1}$Jodrell Bank Centre for Astrophysics, Department of Physics and Astronomy, The University of Manchester, Oxford Road, Manchester M13 9PL, UK\\
$^{2}$Department of Physics and Astronomy, Macquarie University, Sydney, NSW 2109, Australia\\
$^{3}$Astronomy, Astrophysics and Astrophotonics Research Centre, Macquarie University, Sydney, NSW 2109, Australia
}

\date{Accepted 2025 September 10. Received 2025 September 10; in original form 2025 July 07}

\pubyear{\the\year{}}

\begin{document}
\label{firstpage}
\pagerange{\pageref{firstpage}--\pageref{lastpage}}
\maketitle

\begin{abstract}
Planetary nebulae trace the hottest and most luminous phase of evolution of solar-type stars. We use these hot, bright stars to investigate extinctions towards a complete sample of 262 confirmed PNe with large angular diameters, which have the most reliable photometry and hottest central stars. For 162 of these PNe, we identify central stars, produce spectral energy distributions from survey data using \textsc{PySSED},  then fit reddened model spectra to the observed photometry to obtain extinctions accurate down to $E(B-V)$ of $\pm 0.02$\,mag. 
The fitting is performed by Nelder-Mead $\chi^2$ minimisation, with uncertainties evaluated through MCMC.  
The catalogue of stellar temperatures is updated for our sample for the calculation of luminosities.
The extinctions agree well with interstellar extinction. We find evidence of circumnebular extinction for one PN, and evaluate its effect on the planetary nebulae luminosity function. Four new close binaries are identified from the spectral energy distributions. The binary fraction in the full sample is between 23\%\ and 36\%. We use our compiled data to evaluate the quality of the central star identifications in the literature. Three objects in our sample have previously been classified as post-RGB systems but we find that their parameters may also be consistent with post-AGB evolution.

\end{abstract}




\defcitealias{2016MNRAS.455.1459F}{FPJ16}
\defcitealias{1999PASP..111...63F}{F99}
\defcitealias{1989ApJ...345..245C}{CCM89}
\defcitealias{2021A&A...656A.110C}{CW21}
\defcitealias{2021A&A...656A..51G}{GS21}

\section{Introduction}

Planetary nebulae (PNe) are a brief phase in the late stellar evolution of low- and intermediate-mass stars. They are important tracers of stellar evolution and play an important role in the chemical evolution of galaxies \citep{2022PASP..134b2001K}.  They are also used as tracers of stellar populations in distant galaxies, and the planetary nebula luminosity function (PNLF) has become a standard candle for distances \citep{2018NatAs...2..580G, 2025ApJ...983..129J}. PNe are present in both young and old stellar populations, and form the highest luminosity phase of the evolution of these stars, observable to large distances.

Extinction of PNe is caused by both interstellar and local dust. It is most commonly measured from Balmer line ratios. Extinction determinations in the literature have been acknowledged to be widely scattered, e.g., by \citet{2016MNRAS.455.1459F} (hereafter \citetalias{2016MNRAS.455.1459F}). This scatter arises from differences in methodology and observational limitations, as well as the choice of extinction law adopted \citep{2016PASA...33...24N, 2004MNRAS.353..796R}.

Accurate extinction determinations are necessary for proper characterisation of PNe. Constraints on extinction allow the luminosity of both the nebula and the central star to be accurately determined \citepalias{2016MNRAS.455.1459F}. Inaccurate luminosity determinations can lead to a cascade of other issues: a wrong progenitor mass leads to flawed predictions for the evolutionary history and chemistry of a PN. Accurate extinctions also allow the spectral line strengths of PNe to be compared over a broad range of wavelengths, required for accurate abundance determinations. 

PNe are present throughout the Galaxy and are sufficiently luminous to be studied to large distances. They provide ideal tracers of interstellar extinctions along all lines of sight \citep{2016PASA...33...24N}. The stellar spectral energy distribution (SED) can, in principle, provide information on the extinction law in different Galactic environments \citep{2013A&A...550A..35P} where other tracers are sparse or unavailable.

This paper makes use of a new set of central stars of planetary nebulae (CSPNe) identifications made possible by \emph{Gaia} (\citet{2021A&A...656A.110C}, hereafter \citetalias{2021A&A...656A.110C}; \citet{2021A&A...656A..51G}, hereafter \citetalias{2021A&A...656A..51G}). Stellar SEDs are generated for the CSPNe using \textsc{PySSED} \citep{2024RASTI...3...89M}.  For CSPNe with temperatures $> 60$\,kK the spectral shape has only a weak dependence on temperature from the far-UV (1500\AA) through the near-IR (1$\mu$m). We select a complete sample of extended PNe with minor axis $> 1'$, whose CSPN can be expected to have evolved to these high temperatures and where nebular emission can be more easily separated from  the SED of the CSPN. We then fit the SED with an extincted model atmosphere to extract the line-of-sight extinction towards the CSPN. 

The paper is structured as follows. Section 2 compares methods to determine extinction. Section 3 describes the SED fitting procedure. The data is described in Section 4. The results are given in Section 5, including individual objects of interest and binaries. Section 6 contains the discussion.

\section{Extinction}

\subsection{Extinction laws}
\label{sec:ext_laws}

Extinction laws approximate the relative extinction as function of wavelength. The extinction laws most commonly used to study PNe are those of \citet{1999PASP..111...63F}, hereafter \citetalias{1999PASP..111...63F}, and \citet{1989ApJ...345..245C}, hereafter \citetalias{1989ApJ...345..245C}. The older law from \citet{1983MNRAS.203..301H}  also remains in common use. Each of these define a family of profiles based on the parameter $R_V = A_V / E(B-V)$.

Values of $R_V$ along different lines of sight vary between 3.0 and 3.5 in the Galaxy \citep{Lallement2024}, while lower values of $R_V$  towards the Galactic Bulge \citep{2004MNRAS.353..796R} remain disputed \citep{2013A&A...550A..35P, 2016PASA...33...24N}. In this study, we use the \citetalias{1999PASP..111...63F} extinction law and adopt the Galactic standard $R_V=3.1$. Sensitivity to $R_V$ is beyond the scope of this work. The \citetalias{1999PASP..111...63F} extinction law generally agrees well with the extinction law of \citetalias{1989ApJ...345..245C}, but 
differs around the wavelength of H$\alpha$.

Extinctions to planetary nebulae are commonly expressed in terms of the colour excess $E(B-V)$ and the logarithmic extinctions to $H\alpha$ and $H\beta$, respectively $C_\alpha$ and $C_\beta$.
The parameters $C_\alpha$, $C_\beta$ and $E(B-V)$ are related by the extinction law used. We calculated the conversion between the three parameters for the extinction laws of \citetalias{1999PASP..111...63F} and \citetalias{1989ApJ...345..245C}, using the {\tt dust\_extinction} Python package of \citet{2024JOSS....9.7023G}. The results are shown in Table \ref{tab:ExtLaws}, and show that $C_\beta$ relates more consistently to $E(B-V)$. 

\begin{table*}
\centering
\caption{Relation between extinction parameters, for $R_V=3.1$ and 2.5. $B$ and $D$ are defined in Eqs. (\ref{eq:C_alpha_opt}) and (\ref{eq:EBV}), respectively. By definition, $B_\beta$ is always one more than $B_\alpha$. The predicted observed [{F({\rm H}$\alpha$)/F({\rm H}$\beta$)}]$_{\rm obs}$ assumes an intrinsic Balmer ratio of 2.85. Note: the \citet{1983MNRAS.203..301H} extinction law does not have a relation to $R_V$ at wavelengths longer than $5460$\,\AA, where it is defined assuming $R_V=3.1$. An offset of $R_V-3.1$ is applied to the infrared portion of the law to prevent discontinuities at $5460$\,\AA.}
\label{tab:ExtLaws}
\begin{tabular}{ll@{\hspace{30pt}}lll@{\hspace{30pt}}clllc}
\toprule
& & & & & & \multicolumn{4}{c}{$E(B-V)=1$} \\
Extinction law & $R_V$ & $B_\alpha$ & $B_\beta$ & D &   & $A_{5500}$ & $C_\alpha$ & $C_\beta$ & $\left[F({\rm H}\alpha)/F({\rm H}\beta)\right]_{\rm obs}$ \\
\midrule
\citetalias{1989ApJ...345..245C} & 3.5 & 2.60 & 3.60 & 2.24 & & 3.50 & 1.16 & 1.61 & 9.26 \\   
\citetalias{1999PASP..111...63F}  & 3.5 & 2.14 & 3.14 & 1.95 &  & 3.43 & 1.09 & 1.61 & 8.83 \\ 
\citet{1983MNRAS.203..301H} & 3.5 & 2.47$\dagger$ & 3.47$\dagger$ & 2.15$\dagger$ &  & 3.48$\dagger$ & 1.15$\dagger$ & 1.61 & 8.31$\dagger$ \\
\citetalias{1989ApJ...345..245C} & 3.1 & 2.36 & 3.36 & 2.33 &  & 3.10 & 1.01 & 1.44 & 7.67 \\   
\citetalias{1999PASP..111...63F}  & 3.1 & 1.87 & 2.87 & 1.99 &  & 3.03 & 0.94 & 1.45 & 9.08 \\ 
\citet{1983MNRAS.203..301H} & 3.1 & 2.12 & 3.12 & 2.15 &  & 3.08 & 0.99 & 1.45 & 8.31 \\
\citetalias{1989ApJ...345..245C} & 2.5 & 1.97 & 2.97 & 2.04 &  & 2.50 & 0.80 & 1.20 & 7.23 \\   
\citetalias{1999PASP..111...63F}  & 2.5 & 1.46 & 2.46 & 2.04 &  & 2.43 & 0.72 & 1.21 & 8.83 \\ 
\citet{1983MNRAS.203..301H} & 2.5 & 1.61$\dagger$ & 2.61$\dagger$ & 2.15$\dagger$ &  & 2.48$\dagger$ & 0.75$\dagger$ & 1.21 & 8.31$\dagger$ \\
\bottomrule
\end{tabular}
\newline $\dagger$These values are based on the wavelength region $>5460$\,\AA\, where $R_V=3.1$ is assumed by \citet{1983MNRAS.203..301H}.
\end{table*}

\citet{2013MNRAS.431....2F} and \citetalias{2016MNRAS.455.1459F} use the extinction law of \citet{1983MNRAS.203..301H}. 
\citetalias{2016MNRAS.455.1459F} convert observed H$\alpha$/H$\beta$ ratios to $E(B-V)$ but provide neither the original ratios nor the conversion factors.

\subsection{Extinction Measures}

\subsubsection{Balmer decrement}

PNe extinctions are most frequently characterised by their effect on the H$\alpha$ and H$\beta$ flux ratio,  $F({\rm H}\alpha)$/$F({\rm H}\beta)$. 
The intrinsic ratio can vary between 2.75 and 3.0 for $T_{\rm e}= [5\times10^3, \, 2\times10^4]$\,K, corresponding to a change in $C_\beta$ of $\pm 0.03$.
With sufficient spectroscopic data, the intrinsic F(H$\alpha$)/F(H$\beta$) ratio may be determined from detailed photoionization models \citep{2021PASP..133i3002U} but this is not available for the majority of PNe. We adopt an intrinsic ratio of $(F({\rm H}\alpha)$/$F({\rm H}\beta))_0=2.85$ which assume a case-B (low density) nebula with typical electron density of $n_{\rm e}=10^4$\,cm$^{-3}$ and electron temperature $T_{\rm e}=10^4$\,K. The relations become

\begin{equation}
\label{eq:C_alpha_opt}
    C_{\alpha , \text{opt}} = B_\alpha \log \left[ \frac{F({\rm H}\alpha)}{2.85 F({\rm H}\beta)}\right]; \quad C_{\beta , \text{opt}} = B_\beta \log \left[ \frac{F({\rm H}\alpha)}{2.85F({\rm H}\beta)}\right],
\end{equation}

\noindent where the values of $B_\alpha$ and $B_\beta$ for the different extinction laws are given in Table \ref{tab:ExtLaws}.

The relation between $E(B-V)$ and the Balmer line ratio is

\begin{equation} 
\label{eq:EBV}
E(B-V) = D \log\left[\frac{F(H\alpha)}{2.85 F(H\beta)} \right] = D \log\left[\frac{F(H\alpha)}{F(H\beta)} \right] - 0.455. 
\end{equation}

\noindent Values for $D$  are shown in Table \ref{tab:ExtLaws}.

Table \ref{tab:ExtLaws} shows that the value of $B_\alpha$  
differs by 20\%\ between \citetalias{1989ApJ...345..245C} and \citetalias{1999PASP..111...63F}. \citet{2004MNRAS.353..796R} find $B_\alpha =  2.28$ based on the older extinction curve from \citet{1984ASSL..107.....P}. The parameter $D$ can differ by 15\%\ based on the choice of extinction law.
It is undetermined which extinction law yields more accurate $B$ and $D$ values.

\subsubsection{CSPNe extinctions}
\label{sec:cspn_exts}

Extinction to a PN may be determined by evaluating $E(B-V)$ to the central star. CSPNe are very hot, and therefore have a well-determined intrinsic colour at optical wavelengths. \citet{1984ASSL..107.....P} uses an intrinsic colour $(B-V)_0 = -0.33$.
In papers by \citet{2013MNRAS.428.2118D} and \citet{2015MNRAS.448.3132D}, literature CSPNe temperature determinations are used to predict $(B-V)$ colours. This method determines $E(B-V)$; a dust extinction law is used to obtain the total extinction. 
This is the approach that this paper employs, using photometry over a much wider range of wavelengths. Further detail is given in the following Sec.~\ref{sec:methodology}.

\subsubsection{Radio--optical extinctions}
\label{sec:rad_opt_ext}

Dust has a minimal effect longward of the infrared, so observed radio fluxes of planetary nebulae are independent of extinction \citep{1998ApJS..117..361C, 1989A&AS...79..329Z}. The ratio of radio flux over optical hydrogen line flux yields $C_\alpha$ or $C_\beta$ \citep{1984ASSL..107.....P}, independent of the extinction law.

The method has been applied  by \citet{1998ApJS..117..361C} and \citet{2004MNRAS.353..796R}. There has been a tendency for radio extinctions to be higher than optical ones \citep[e.g.,][]{2013A&A...550A..35P}, and it is unknown whether this is a matter of observational uncertainties or physics assumptions.  It is best used for compact PNe, as the radio brightness temperature drops dramatically with size and the integrated radio flux becomes difficult to accurately measure.

\subsubsection{Galactic tomographic extinctions}

Three-dimensional dust maps of the local Milky Way have become available based on \emph{Gaia} stellar parallaxes. If the distance to a nebula is known, then these dust maps, together with an extinction law, can be used to predict an interstellar extinction. Two-dimensional dust maps can be used for objects outside the Galactic Plane, ergo beyond the dust extinction column of the disk.

This paper uses the \textsc{G-Tomo} (Galactic tomography) dust map \citep{2022AnA...661A.147L, 2022yCat..36640174V}, which provides extinction estimates to regions up to 6\,kpc from the Sun based on stars with good \emph{Gaia} and 2MASS photometry and with accurate parallaxes. 

This is a powerful method if the  extinction cube are accurate, which requires a high stellar density along a line of sight. The method can  miss small-scale structure in the interstellar medium. It also becomes less reliable at large distance or high extinction. 

The distance to the PN is also needed. 
The advent of \emph{Gaia} has revolutionised distance measurement for PNe whose central star can be identified \citep{2021A&A...656A..51G, 2021A&A...656A.110C}. 
In the absence of CSPNe parallaxes, the most rigorous statistical distance method for PNe is the $F({\rm H}\alpha)$ surface brightness-radius distance indicator described by \citetalias{2016MNRAS.455.1459F}. Other approaches to primary and statistical distance measurements of PNe are discussed in \citet{2008PhDT.......109F}.

\section{Methodology}
\label{sec:methodology}

A wide range of photometric measurements is used of the CSPNe to derive the stellar SED and fit the $E(B-V)$ using stellar atmosphere models.

\subsection{PySSED}
\label{sec:pyssed}

We have used \textsc{PySSED} v1.1.20241206\footnote{\url{https://github.com/iain-mcdonald/PySSED}} \citep{2024RASTI...3...89M, 2025MNRAS.tmp..943M} for the selected CSPNe. \textsc{PySSED}  generates SEDs based on archival photometric and distance data and stellar atmospheres to derive stellar temperature and luminosity. The SEDs are obtained from  broadband stellar photometry in selected VizieR\footnote{\url{https://vizier.cds.unistra.fr}} catalogues  at CDS, Strasbourg. \textsc{PySSED} merges the data, performs outlier rejection and fits \textsc{BT-Settl} atmosphere models, appropriate for most typical stars. A comprehensive description of \textsc{PySSED} is given in \citet{2024RASTI...3...89M}.

We adapted \textsc{PySSED} for this study of CSPNe by including appropriate atmosphere models and performing extinction fitting. The differences in the fitting procedures from the base version of \textsc{PySSED} are described in Section \ref{sec:ext_det_method}. 
For distances to stars, \textsc{PySSED} prioritises \emph{Gaia} DR3 distance from \citet{2021AJ....161..147B}.
These distances are combined with other available distance measurements, e.g. from \emph{Hipparcos} parallaxes, in an error-weighted average. In this work we obtain distance uncertainties from the uncertainties of the sources using the same error-weighted average. 

\subsection{CSPN atmosphere models}
\label{sec:limitations}

CSPNe evolve on post-AGB evolutionary tracks. The stars evolve at constant luminosity of typically $5\times10^3$--$10^4$\,L$_\odot$ from the AGB turn-off to temperatures above $10^5$\,K, after which stars enter the high-temperature end of the white-dwarf cooling track \citep{2016A&A...588A..25M}. In this study, TMAP atmosphere models are adopted; these are a suite of non-LTE models which describe the spectra of compact hot stars with different temperatures, surface gravities and compositions \citep{1999JCoAM.109...65W, 2003ASPC..288...31W, 2003ASPC..288..103R}. The TMAP models were retrieved from the TheoSSA\footnote{\url{http://dc.g-vo.org/theossa}} web service \citep{vo:tmap_web, 2018MNRAS.475.3896R}.

The filters used here range in wavelength from \emph{GALEX} FUV (1549\AA) to Pan-STARRS $Y$ (9613\AA). The extinction is derived under the  assumption that the intrinsic shapes of CSPNe spectra do not show strong variation over this wavelength range. This assumption is tested using TMAP models by examine the variation of these spectra in the temperature range $20 - 150$\,kK at UV through IR wavelengths. 
 
\begin{figure*}
\includegraphics[width=\textwidth]{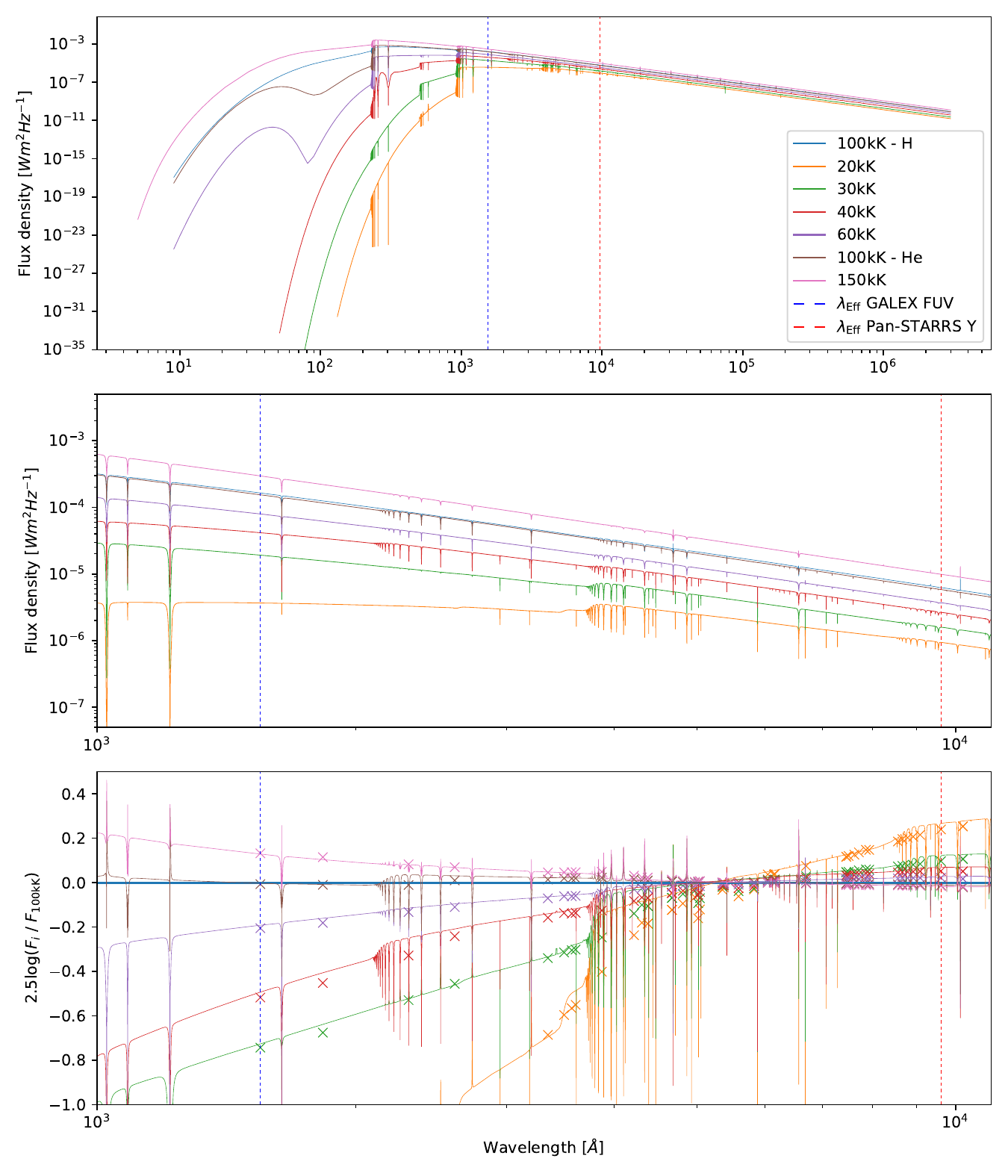}
\caption{The spectra of CSPNe  according to the TMAP models. The top panel shows the full breadth of the TMAP spectra, while the middle panel focuses on the 1000\AA\ to 10,000\AA\ range that is of interest to this investigation. The bottom panel shows how these spectra vary relative to the 100kK model in magnitudes, normalised at the Johnson $V-$band wavelength.  
The crosses represent the convolutions of these spectra with the filters in which CSPNe are commonly observed,  plotted at the effective wavelength of the filter}. 
\label{fig:TempComp}
\end{figure*}

Figure \ref{fig:TempComp} shows the TMAP CSPNe spectra at different temperatures. The bottom panel plots  the models  normalised at the Johnson $V-$band relative to the 100\,kK model. When normalised, the 60\,kK and 150\,kK models are within 0.21\,mag of the 100\,kK model for all the filters we consider: the discrepancy of the 60\,kK model ranges from $-0.21$\,mag at \emph{GALEX} FUV to $+0.02$\,mag at Pan-STARRS $Y$, while the discrepancy of the 150\,kK model ranges from $+0.13$\,mag to $-0.02$\,mag over the same filters. The largest variation exists at UV wavelengths, where the spectra are nearing the peak of their intrinsic SED. 
At optical wavelengths $>4000$\,\AA, the variation is less than 0.1\,mag for temperatures down to 40\,kK. The 20\,kK spectrum has a significantly different shape, with a strong Balmer jump.

Figure \ref{fig:TempComp} demonstrates that the composition of CSPNe is not an important factor within the wavelength range that we consider. The hydrogen and the helium dominated models at $10^5$\,K are negligibly different, with discrepancies $<0.02$\,mag for all the filters we consider. The difference at shorter wavelengths $<400$\,\AA\, leads to a $1.7\%$ difference in predicted bolometric luminosity between the hydrogen and helium dominated cases - negligible compared with the overall luminosity uncertainty, see Sec.~\ref{sec:luminosities}.  We tested the impact of varying the surface gravity and found a negligible effect, so we do not consider this parameter further.

Based on this result, we adopt as the standard spectral shape for CSPNe the TMAP model at $T=10^5$\,K, $\log(g)=6$, $\text{H}=0.9$, $\text{He}=0.1$, which is taken as a good approximation for any CSPNe with $T \ge 60\,000$ K, as will be the case for CSPNe on the cooling track. 
The effect of this assumption on extinction results depends on the data available in a specific SED, but is typically of order $E(B-V)\approx 0.02$\,mag for the $60\,000$\,K and $150\,000$\,K cases relative to $10^5$\,K. Further detail on the uncertainty handling is given in Sec.~\ref{sec:uncertainties}.

\subsection{SSED extinction determination}
\label{sec:ext_det_method}

Where previous works have used single colours to determine CSPNe extinctions \citep{2013MNRAS.428.2118D}, here we draw upon a wider range of photometry to fit an extinction to the whole SED of the star. 

Caution is required when considering the effect of extinction on an SED. Many of the photometric data come from observations through broadband filters in surveys such as \emph{Gaia}. The wavelength dependence of extinction can introduce a strong, potentially non-linear colour term over the wavelength range of a filter, depending on the photon spectrum. In the generic case, this spectrum may not be known, making extinction difficult to correct for in observations. In the case of CSPNe, the photon distribution is represented by an adopted standard spectral shape.

We determine extinction by comparing the observed SED for the CSPN, created by \textsc{PySSED}, to a set of reddened model SEDs, based on the adopted TMAP model. The model SEDs are reddened  according to the dust extinction law of \citetalias{1999PASP..111...63F}, quantified by  the colour excess $E(B-V)$ and $R_V=3.1 $ used as standard for the Milky Way. We use the Python package \textsc{dust\_extinction} to handle extinction \citep{2024JOSS....9.7023G}. The reddened spectrum is then convolved with the filter-transmission curves obtained from the Spanish Visual Observatory Filter Profile Service\footnote{\url{http://svo2.cab.inta-csic.es/svo/theory/fps3/index.php}} \citep{2012ivoa.rept.1015R, 2020sea..confE.182R} to find a model SED of synthetic flux observations. We produce reddened SEDs in the range $E(B-V)=-0.5$ to $E(B-V)=3$. By allowing negative extinctions as potential solutions we obtain a gauge on the accuracy.

To normalize the reddened model to  the observed SED, we calculate for each filter the flux difference between model and observed flux.  The median value of these differences is applied as an offset to align the model SED. 
The Nelder--Mead algorithm is used to minimise the reduced $\chi^2_{\rm{r}}$ with $E(B-V)$ as the fitted parameter. The best fit $E(B-V)$ may be transformed into a $C_\beta$ or $C_\alpha$ value using Table \ref{tab:ExtLaws}.

Photometric data points within the wavelength range 1000\,\AA\,to 10\,000\,\AA\, are used for these fits. The lower end corresponds to the shortest wavelength for which the \citetalias{1999PASP..111...63F} dust extinction law is defined. At wavelengths greater than 10,000\,\AA, IR excesses or nebular contamination can begin to have significant impacts on the measured SED, so it cannot consistently be claimed that the observed SEDs are dominated by the CSPNe above this wavelength. 

\subsubsection{Uncertainties}
\label{sec:uncertainties}

Rigorously determining accurate uncertainties for quantities fitted to SEDs created from survey data is challenging (see \citet{2024RASTI...3...89M}). These SEDs constitute a heterogeneous dataset with different uncertainties. Issues may include erroneous cross-matching, photometric contamination, improper calibration, etc. \textsc{PySSED} assumes a minimum uncertainty for each survey which may be larger than the reported calibration uncertainty \citep{2024RASTI...3...89M}.

We have used  two approaches to find the accuracy of our extinctions. The first is to evaluate the uncertainty suggested by the spread of the data. We do this using the \textsc{emcee} Python package \citep{2013ascl.soft03002F} to perform Markov-Chain--Monte-Carlo (MCMC) analysis on the space of possible $E(B-V)$ solutions, using the reduced $\chi^2_{\rm{r}}$ as the goodness-of-fit measure. We set an uncertainty floor for each star based on the minimum of the reduced $\chi^2_{\rm{r}}$  and for each data point, adopt the higher of the uncertainty floor or the uncertainty quoted by its survey. We bootstrap uncertainties by performing an initial test of the data with an uncertainty floor of 5\% on our data points, with 100 walkers performing 1000 walks, then adjust the uncertainty floor by the inverse square root of the reduced $\chi^2_{\rm{r}}$ of this fit. This ensures that the final fit will have a reduced $\chi^2_{\rm{r}}$ close to unity. A longer MCMC fit with 10\,000 walks is then performed with the new uncertainty floor to find estimates for the upper and lower reasonable extinctions. We adopt these as uncertainties on our extinction result. This MCMC analysis also provides reassurance that the solutions of the Nelder--Mead optimisation are not local minima.

Second, we evaluate any bias on our extinction introduced from the implicit assumption that the star has an effective temperature of $100\,000$\,K. To evaluate this error, we repeat the standard analysis but instead with TMAP models at both $T=60,000$\,K and $T=150,000$\,K. 
The result of these fits gives another set of upper and lower extinction values, which are taken to represent a reasonable estimate of the uncertainty introduced by assuming a fixed temperature. The size of these temperature derived uncertainties vary based on the data in the SED but are typically of order $-0.02$\,mag relative to the $T=60\,000$\,K model and $+0.02$\,mag relative to the $T=150\,000$\,K when measured as $E(B-V)$.

The uncertainties from the two approaches are added in quadrature to find an overall uncertainty on our extinction results. This uncertainty is not a formal $1\sigma$ determination, as that is challenging to evaluate for this type of data, but it serves as a useful approximant.

\section{Data}

\subsection{Sample selection and CSPN candidate evaluation}
\label{sec:cspn_grades}

A complete sample of extended planetary nebula was selected from the HASH database \citep{2016JPhCS.728c2008P}. The sample criteria were that a source should have a {\tt True} PN status within HASH and that it should have a minor axis greater than one arcminute. A total of 262 detected sources in HASH meet these criteria.

The choice to focus on large PNe is motivated by several factors. Compared to compact PNe, this population has generally reduced nebular emission in the vicinity of their CSPN, which means minimal contamination of the CSPN SED. We can also expect the intra-nebular extinction of these PNe to be minimal, so our measured extinctions can be perceived as mostly interstellar. Additionally, large PNe are more evolved objects whose central stars will more likely fulfil our methodology's assumption of hot CSPN temperatures.

The minor axis is used to avoid cases where the major axis is dominated by bipolar lobes. The angular size is used over physical size because distances are not available for the complete sample, and because of its relevance to  avoiding nebular contamination of the CSPN photometry. Considering physical sizes of sources with reliable distance determinations, the smallest one in our sample, NGC 6720, has a minor axis of $0.25$\,pc, i.e. is still a large, evolved PN. The largest, Sh 2-176, has a minor axis of 3.9\,pc.

The CSPN identifications build upon the work of \citetalias{2021A&A...656A.110C} and \citetalias{2021A&A...656A..51G}, who have used 
\emph{Gaia} to identify candidate CSPN for many Galactic PNe. For some sources, particularly more recent PN discoveries, we have manually identified further CSPN candidates. In total, we were able to obtain CSPN candidates for 255 out of 262 PNe.

The identifications of \citetalias{2021A&A...656A.110C} and \citetalias{2021A&A...656A..51G} were produced algorithmically based on the star's position in the sky relative to the nebula and its \emph{Gaia} $G_\text{BP}-G_\text{RP}$ colour. This approach is broad and useful, but is acknowledged to lead to some misidentifications. Both \citetalias{2021A&A...656A.110C} and \citetalias{2021A&A...656A..51G} provide quality indicators to indicate their confidence in particular identifications.

The data collated in this work provide an opportunity to more deeply evaluate the candidature of the CSPNe within our sample. For each \emph{Gaia} source, we 
evaluate the likelihood of a correct CSPN identification using the available information. For a CSPN identification to be considered reliable, we require that the luminosity derived from the 100\,kK TMAP model, $L_\text{100kK}$, is consistent with the expected range $10-10\,000\,\text{L}_\odot$; and we require that the SED is consistent with the emission of a CSPN. This second criterion can be uncertain, as for stars with fewer data it can be difficult to distinguish a main sequence star from a highly extincted CSPN based on the SED alone. We therefore also consider a range of auxiliary criteria, which include: the derived $E(B-V)$ compared to other extinction indicators for the nebula, primarily those from \textsc{G-Tomo} and \citetalias{2016MNRAS.455.1459F}; our luminosity result for the CSPN, the comparison of our luminosities to those compiled by \citet{2020A&A...640A..10W}
; the \textit{Gaia} derived distance for the CSPN candidate compared to the PN distance found by \citetalias{2016MNRAS.455.1459F}; any information on the PN and CSPN included in HASH \citep{2016JPhCS.728c2008P}; as well as a visual inspection of the source and CSPN position using images available in HASH and the Aladin Sky Atlas\footnote{\url{https://aladin.cds.unistra.fr}} \citep{2022ASPC..532....7B}. Detection of UV emission in the location of the CSPN candidate, even if not accurate enough to include in the SED, is taken as a positive indicator of CSPN status.
CSPNe not detected by \emph{Gaia} are not included.

The 262 sources with their grades are given in the online supplement to the paper; a truncated set of grades are shown in Table \ref{tab:cspn_grades}. We use the sum of evidence available for each source to grade our CSPN candidates. 
An A grade is given to sources whose luminosities are consistent with the range found in the evolutionary tracks of \citet{2016A&A...588A..25M} and with an SED which is consistent with being a CSPN; an A+ grade is given to an A candidate with auxiliary evidence that clearly agrees with CSPN status. 
A B grade is given to candidates with ambiguous SEDs, where there are few data points and when auxiliary information is not conclusive; while a C grade is given where the auxiliary information suggests the source is unlikely to be a CSPN. An F grade is given to stars which are found to unambiguously be CSPN misidentifications, i.e. whose luminosities or SEDs are inconsistent with CSPN status. A U grade is given to stars for which there are insufficient data to make any assessment or where no CSPN is identifiable. There remains an element of qualitative judgement regarding these grades.

\begin{table*}
\centering
\caption{Grades indicating the reliability of the CSPN identifications. The reliability indicators of \citetalias{2021A&A...656A.110C} and \citetalias{2021A&A...656A..51G} are included for reference. The grading criteria used by this work is described in Sec.~\ref{sec:cspn_grades}. The full table is available on-line.}
\label{tab:cspn_grades}
\begin{tabular}{llllccll}
\toprule
PN G & PN Name & CSPN \emph{Gaia} ID & \multicolumn{1}{p{1cm}}{\centering Reliability \\ Grade} & \multicolumn{1}{p{1cm}}{\centering CW21 \\ Reliability \\ Score} & \multicolumn{1}{p{1cm}}{\centering GS21 \\ Reliability \\ Group} & Distance$^a$ [pc] & Frew Distance$^b$ [pc] \\
\midrule
002.7-52.4 & IC 5148/50 & 6574225217863069056 & A+ & 1.00 & A & $1170 \pm 70$ & $1400 \pm 400$ \\
013.8-02.8 & SaWe 3 & 4096221021593115264 & C & 0.99 & B & $1900 \pm 300$ & $2000 \pm 700$ \\
017.3-21.9 & Abell 65 & 6864617817991978624 & A & 1.00 & B & $1420 \pm 70$ & $1400 \pm 300$ \\
019.4-19.6 & K 2-7 & 6868431267910764160 & A+ & 1.00 & A &  & $2200 \pm 400$ \\
019.8-23.7 & Abell 66 & 6864496115795567488 & A+ & 1.00 & A & $1200 \pm 200$ & $1100 \pm 300$ \\
025.0-11.6 & Abell 60 & 4186077574260953984 & A+ & 1.00 & A &  & $2900 \pm 600$ \\
025.4-04.7 & IC 1295 & 4203649492421426816 & A & 0.98 & A & $1600 \pm 200$ & $1100 \pm 200$ \\
033.1-06.3 & NGC 6772 & 4261038467432411776 & A & 0.97 & B & $1000 \pm 200$ & $1300 \pm 400$ \\
034.1-10.5 & HaWe 13 & 4213334746695221888 & A+ & 0.99 & A & $1800 \pm 200$ & $2600 \pm 500$ \\
035.9-01.1 & Sh 2-71 & 4268419317167114240 & A & 1.00 & A &  & $1500 \pm 500$ \\
036.0+17.6 & Abell 43 & 4488953930631143168 & A+ & 1.00 & A & $2090 \pm 110$ & $2400 \pm 500$ \\
036.1-57.1 & NGC 7293 & 6628874205642084224 & A+ & 1.00 & A & $199 \pm 2$ & $240 \pm 70$ \\
037.9-03.4 & Abell 56 & 4268179207028750592 & A & 1.00 & A &  & $2100 \pm 600$ \\
038.7+01.9 & YM 16 & 4282704172233945216 & F &  & A &  & $1200 \pm 400$ \\
041.8-02.9 & NGC 6781 & 4294123077230164736 & A & 1.00 & B & $494 \pm 19$ & $800 \pm 200$ \\
043.5-13.4 & Abell 67 & 4241183550857632896 & A+ & 1.00 & A & $3800 \pm 1300$ & $4100 \pm 1200$ \\
044.3+10.4 & We 3-1 & 4509807233807831936 & A*OB & 1.00 & B & $3700 \pm 1000$ & $1900 \pm 600$ \\
046.8+03.8 & Sh 2-78 & 4506484097383382272 & A+ & 1.00 & A & $700 \pm 50$ & $800 \pm 200$ \\
047.0+42.4 & Abell 39 & 1305573511415857536 & A+ & 1.00 & A & $1160 \pm 50$ & $1700 \pm 300$ \\
047.1-04.2 & Abell 62 & 4314617943916816384 & B & 1.00 & A & $6400 \pm 2600$ & $1600 \pm 400$ \\
053.3+03.0 & Abell 59 & 4516549404740191488 & A & 0.90 &  &  & $2100 \pm 800$ \\
055.4+16.0 & Abell 46 & 4585381817643702528 & A+ & 0.99 & A & $2500 \pm 200$ & $2100 \pm 400$ \\
059.7-18.7 & Abell 72 & 1761341417799128320 & A+ & 1.00 & A & $1840 \pm 200$ & $2000 \pm 400$ \\
060.8-03.6 & NGC 6853 & 1827256624493300096 & A+ & 1.00 & A & $387 \pm 7$ & $350 \pm 100$ \\
063.1+13.9 & NGC 6720 & 2090486618786534784 & A+ & 1.00 & A & $780 \pm 30$ & $1000 \pm 300$ \\
$\vdots$ & $\vdots$ & $\vdots$ & $\vdots$ & $\vdots$ & $\vdots$ & $\vdots$ & $\vdots$ \\
\bottomrule
\end{tabular}

    *The stars denoted *OB are fit as optical binaries, see Sec.~\ref{sec:composite_seds}. $^a$The distance derived by \textsc{PySSED} from \emph{Gaia / Hipparcos} parallax, only quoted for stars with fractional parallax errors $< 0.5$. $^b$Refers to the surface brightness distances of \citetalias{2016MNRAS.455.1459F}.

\end{table*}

\subsection{Survey data}

The survey data used to produce the SEDs are drawn from the default catalogues available in \textsc{PySSED}, with the exception of SDSS16 which is excluded as its CSPN photometry is frequently inconsistent with the results of other surveys. We have additionally included photometry from the catalogue of PNe detected by \emph{GALEX} compiled by \citet{2023ApJS..266...34G}, which adds UV photometry for some sources not included in the main \emph{GALEX} catalogue, as well as the photometry of \citet{2013MNRAS.428.2118D} which consists of Johnson--Cousins $BVRI$ measurements.

As \emph{Gaia} is the primary catalogue, all stars have \emph{Gaia} photometry. Additional CSPNe photometry is most commonly drawn from \emph{GALEX}, PanStarrs and APASS. 

We mask \emph{GALEX} or \emph{GALEX}-derived data where the flux $ F > 10^{-2.25}$\,Jy, as those UV data points are often in disagreement with the rest of the SED for the brightest stars. This may be due to a saturation effect in the \emph{GALEX} data.  

\section{Results}

\subsection{Extinction fitting} 
\label{sec:ebv_fitting}

\begin{figure}
{\includegraphics[width=\columnwidth]{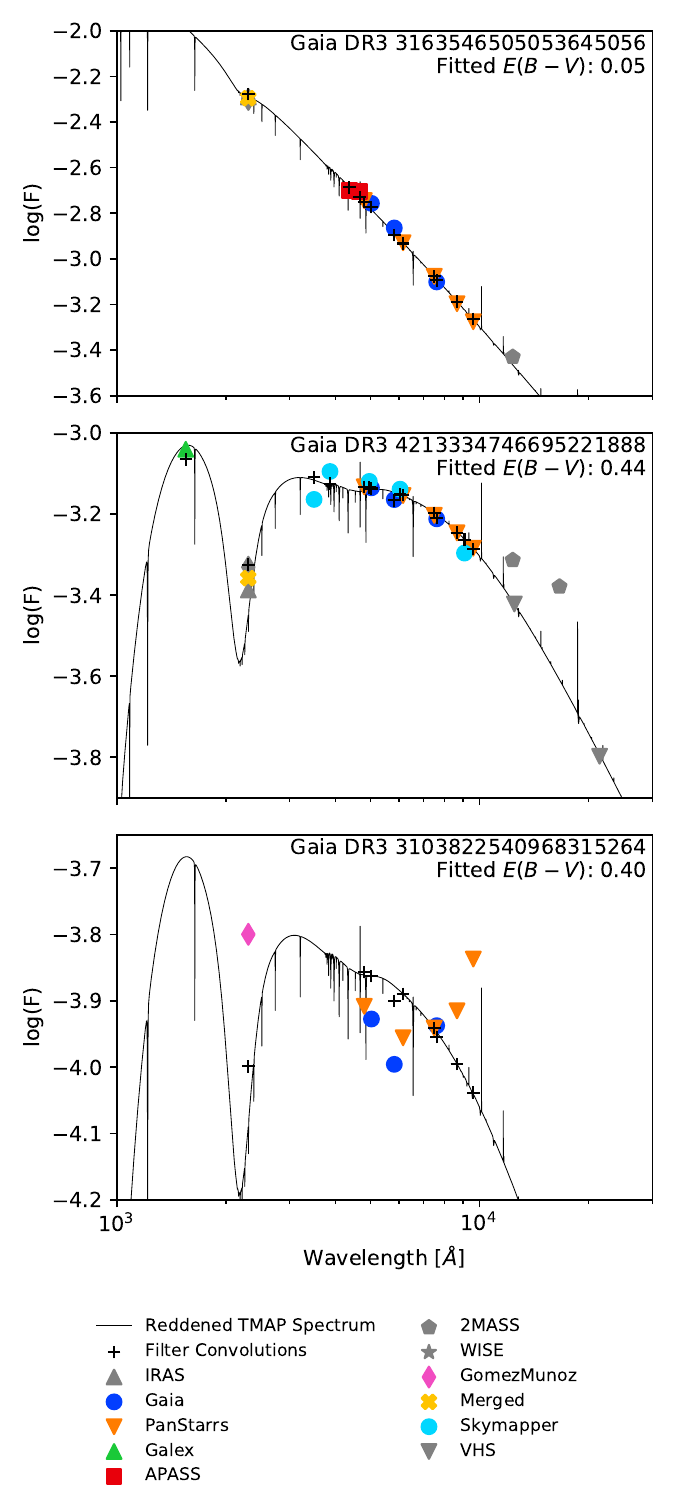}}
\caption{Two successful fits and one unsuccessful fit from the fitting procedure described in Section \ref{sec:ext_det_method}. The shown CSPNe are of Abell 21, HaWe 13 and BMP J0642-0417, respectively. The spectrum is the TMAP model spectrum reddened by the amount of the best fit extinction; the black crosses are the filter convolutions with that spectrum. The coloured markers are observations which are included in the fit, while grey markers which do not have crosses fit to them are data points which are excluded. } 
\label{fig:SuccessfulFits}
\end{figure}

The method is able to produce very good fits between the observations and the models in many cases. Figure \ref{fig:SuccessfulFits} shows examples of successful fits for two CSPNe and an additional case of an unsuccessful fit. The top panel shows the SED for the central star of Abell 21, where the fit finds a low extinction. The middle panel shows the central star of HaWe 13, which has a higher extinction. The 2175\,\AA\, bump in the Galactic extinction profile causes the extinction at the \emph{GALEX} near-UV point to be greater than the extinction at the \emph{GALEX} far-UV point, consistent with the data.

Data points beyond 1\,$\mu$m are excluded from our fitting procedure to avoid nebular contamination and because these points can be affected by main-sequence binary companions or circumstellar dust disks.
The behaviour of these infrared data points is notable. 
For HaWe 13, the 2MASS points are much brighter than the fit predicts, but the VHS points at the same wavelengths fit well. This indicates the importance of the aperture: the 2MASS data points are derived from a large aperture which may have included nebular light. 
Surveys with a larger aperture may also be more prone to measurement errors due to unrelated field stars contained in the photometric aperture.

On the other hand, the bottom panel in Figure \ref{fig:SuccessfulFits} shows an example of a poor fit for the CSPN of BMP J0642-0417. Based on position, UV emission, distance and imaging of the nebula, we are confident that this CSPN candidate is correctly identified, but the photometry is not well reproduced. The \emph{GALEX} data point indicates that the fitted extinction is too high. The SED is suggestive of two stars, one cool and one hot, which cannot be reproduced with a single-star fit. This source demonstrates that some of our sample will be poorly fit, so some quality criteria must be adopted to identify successful fits.  

The results of the extinction fitting for our confident CSPNe identifications (grades A and A+) are given in the online supplement to the paper; a truncated set of results are shown in Table \ref{tab:ExtFits}. This constitutes a set of 162 extinction determinations. The table shows the fitted $E(B-V)$ with uncertainties and $C_\beta$ derived from $E(B-V)$ using the \citetalias{1999PASP..111...63F} extinction law. We present the naive reduced $\chi_r^2$, which is the reduced $\chi_r^2$ found by adopting the non-bootstrapped $5\%$ error floor, as a quality of fit indicator. We consider a fit to be high quality when there are $\ge5$ data points in the SED and the  naive reduced $\chi^2_{\rm{r}} < 5$. Of the 162 extinction fits, 113 meet this criteria. The remaining 49 have larger uncertainties and for sparse SEDs are sensitive to photometric inaccuracies in their few data points. 
For 7 sources,  a fit as binary is presented in Sec.~\ref{sec:composite_seds}, these sources are designated as A*OB in Table \ref{tab:cspn_grades}. 

The \textsc{G-Tomo} extinction is given with uncertainties obtained from the distance uncertainty. Some objects lie beyond the line of sight extinction curve measured by \textsc{G-Tomo}. At higher latitudes ($|b|>10^{\circ}$), the CSPN is then taken to lie beyond the dust column of the Galactic plane and is assigned the asymptotic \textsc{G-Tomo} extinction for that sight line, even if no distance can be obtained, and no uncertainty is provided for the \textsc{G-Tomo} extinction. For CSPN within the Galactic plane ($|b|<10^{\circ}$), the maximum \textsc{G-Tomo} extinction is given as a lower limit.

\begin{table*}
\centering
\caption{The main results of the fitted \textsc{PySSED}  extinctions, given as $E(B-V)$. The full table is available on-line. 
}
\label{tab:ExtFits}
\begin{tabular}{lllllll@{\ }rr}
\toprule
PN G & PN Name & CSPN \emph{Gaia} ID & Distance$^a$ [pc] & \multicolumn{1}{p{1.2cm}}{\centering \textsc{G-Tomo} \\ $E(B\!-\!V)$$^b$} & \multicolumn{1}{p{1.1cm}}{\centering \textsc{PySSED} \\ $E(B\!-\!V)$} & \multicolumn{1}{p{1.1cm}}{\centering \textsc{PySSED} \\ $C_\beta$$^c$} & \multicolumn{1}{p{0.8cm}}{\centering Naive \\ $\chi^2_r$$\,^d$} & SED Points \\ 
\midrule
002.7$-$52.4 & IC 5148/50 & 6574225217863069056 & $1170 \pm 70$ & $0.02$ & $-0.01 \substack{+0.02 \\ -0.06}$ & $-0.02 \substack{+0.02 \\ -0.08}$ & 0.27 & 9 \\
017.3$-$21.9 & Abell 65 & 6864617817991978624 & $1420 \pm 70$ & $0.11$ & $0.30 \substack{+0.02 \\ -0.07}$ & $0.44 \substack{+0.02 \\ -0.10}$ & 45.60 & 18 \\
019.4$-$19.6 & K 2-7 & 6868431267910764160 &  & $0.12$ & $0.17 \substack{+0.02 \\ -0.03}$ & $0.25 \pm 0.04$ & 2.14 & 8 \\
019.8$-$23.7 & Abell 66 & 6864496115795567488 & $1200 \pm 200$ & $0.13$ & $0.22 \substack{+0.02 \\ -0.03}$ & $0.31 \substack{+0.02 \\ -0.04}$ & 1.04 & 10 \\
025.0$-$11.6 & Abell 60 & 4186077574260953984 &  & $0.23$ & $0.14 \substack{+0.01 \\ -0.03}$ & $0.21 \substack{+0.02 \\ -0.05}$ & 2.61 & 14 \\
025.4$-$04.7 & IC 1295 & 4203649492421426816 & $1600 \pm 200$ & $0.47 \pm 0.01$ & $0.20 \substack{+0.03 \\ -0.05}$ & $0.29 \substack{+0.04 \\ -0.07}$ & 3.16 & 9 \\
033.1$-$06.3 & NGC 6772 & 4261038467432411776 & $1000 \pm 200$ & $0.46 \pm 0.01$ & $0.66 \substack{+0.08 \\ -0.09}$ & $0.96 \substack{+0.12 \\ -0.13}$ & 13.53 & 8 \\
034.1$-$10.5 & HaWe 13 & 4213334746695221888 & $1800 \pm 200$ & $0.37$ & $0.44 \substack{+0.01 \\ -0.03}$ & $0.63 \substack{+0.02 \\ -0.05}$ & 1.00 & 15 \\
035.9$-$01.1 & Sh 2-71 & 4268419317167114240 &  &  & $0.78 \substack{+0.10 \\ -0.06}$ & $1.13 \substack{+0.14 \\ -0.09}$ & 7.56 & 6 \\
036.0+17.6 & Abell 43 & 4488953930631143168 & $2090 \pm 110$ & $0.15$ & $0.19 \substack{+0.01 \\ -0.02}$ & $0.28 \substack{+0.02 \\ -0.03}$ & 0.21 & 12 \\
036.1$-$57.1 & NGC 7293 & 6628874205642084224 & $199 \pm 2$ & $0.02$ & $-0.01 \substack{+0.01 \\ -0.02}$ & $-0.02 \substack{+0.02 \\ -0.03}$ & 0.42 & 19 \\
037.9$-$03.4 & Abell 56 & 4268179207028750592 &  &  & $0.28 \substack{+0.06 \\ -0.05}$ & $0.40 \substack{+0.08 \\ -0.07}$ & 3.15 & 6 \\
041.8$-$02.9 & NGC 6781 & 4294123077230164736 & $494 \pm 19$ & $0.31 \pm 0.01$ & $0.54 \substack{+0.02 \\ -0.03}$ & $0.79 \substack{+0.03 \\ -0.04}$ & 1.11 & 8 \\
043.5$-$13.4 & Abell 67 & 4241183550857632896 & $3800 \pm 1300$ & $0.15$ & $-0.03 \substack{+0.02 \\ -0.04}$ & $-0.04 \substack{+0.03 \\ -0.06}$ & 6.74 & 8 \\
046.8+03.8 & Sh 2-78 & 4506484097383382272 & $700 \pm 50$ & $0.32 \substack{+0.02 \\ -0.01}$ & $0.30 \substack{+0.01 \\ -0.02}$ & $0.44 \substack{+0.02 \\ -0.03}$ & 0.12 & 12 \\
047.0+42.4 & Abell 39 & 1305573511415857536 & $1160 \pm 50$ & $0.04$ & $0.07 \pm 0.02$ & $0.11 \pm 0.03$ & 0.99 & 12 \\
053.3+03.0 & Abell 59 & 4516549404740191488 &  &  & $1.17 \substack{+0.07 \\ -0.06}$ & $1.70 \substack{+0.10 \\ -0.08}$ & 2.81 & 5 \\
055.4+16.0 & Abell 46 & 4585381817643702528 & $2500 \pm 200$ & $0.11$ & $0.24 \pm 0.03$ & $0.34 \substack{+0.04 \\ -0.05}$ & 6.50 & 9 \\
059.7$-$18.7 & Abell 72 & 1761341417799128320 & $1840 \pm 200$ & $0.08$ & $0.03 \substack{+0.01 \\ -0.02}$ & $0.04 \pm 0.02$ & 0.25 & 12 \\
060.8$-$03.6 & NGC 6853 & 1827256624493300096 & $387 \pm 7$ & $0.06$ & $-0.00 \substack{+0.00 \\ -0.03}$ & $-0.00 \substack{+0.00 \\ -0.04}$ & 0.16 & 12 \\
063.1+13.9 & NGC 6720 & 2090486618786534784 & $780 \pm 30$ & $0.06$ & $0.00 \substack{+0.07 \\ -0.06}$ & $0.00 \substack{+0.10 \\ -0.09}$ & 5.80 & 12 \\
$\vdots$ & $\vdots$ & $\vdots$ & $\vdots$ & $\vdots$ & $\vdots$ & $\vdots$ & $\vdots$ \\
\bottomrule

\end{tabular}

    $^a$The distance derived by \textsc{PySSED} from \emph{Gaia / Hipparcos} parallax, only quoted for stars with fractional parallax errors  $< 0.5$. $^b$The uncertainties on the \textsc{G-Tomo} $E(B-V)$ are based on the range of the extinction curve bounded by the 1-$\sigma$ distance uncertainty. If the apparent uncertainty is smaller than $0.005$ then it is not given. If no distance is available but the PN lies more than 10$^\circ$ from the Galactic plane, then the \textsc{G-Tomo} extinction refers to the total extinction for that line of sight. $^c$$C_\beta$ is derived using Table \ref{tab:ExtLaws} and \citetalias{1999PASP..111...63F}. $^d$The naive $\chi_r^2$ refers to the reduced $\chi^2$ calculated assuming an error floor of 5\%, i.e. without performing error bootstrapping.

\end{table*}

\begin{figure}
{\includegraphics[width=\columnwidth]{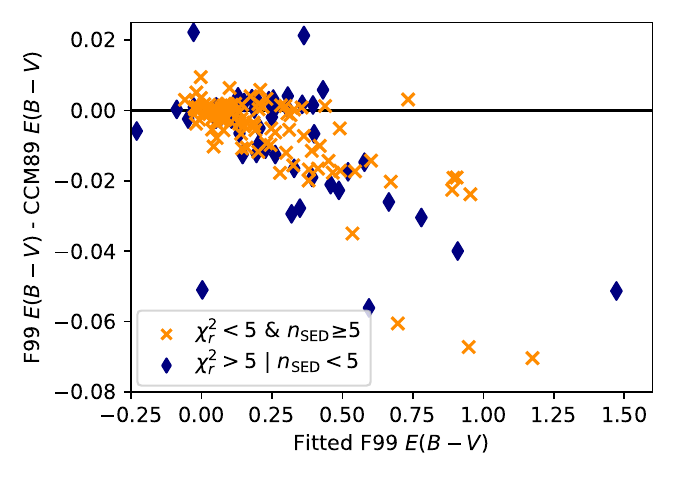}}
\caption{The difference in $E(B-V)$ between adopting the \citetalias{1999PASP..111...63F} or \citetalias{1989ApJ...345..245C} extinction laws. Only CSPNe which have been evaluated with an A+ or A grade are included. High quality fits with naive $\chi^2_r<5$ and $\ge5$ SED points are separated from lower quality fits.}
\label{fig:ext_law_difference}
\end{figure}

To quantify the effect of the adopted extinction law, we refitted the data using the extinction law of \citetalias{1989ApJ...345..245C}. The resulting $E(B-V)$ values were compared with those in Table \ref{tab:ExtFits}, as shown in Figure \ref{fig:ext_law_difference}. 
At low extinction, the results agree in most cases to within $\Delta E(B-V) =0.01$\,mag.  In some cases there are more significant disagreements: this is mainly when the fit is of poorer quality.  At higher extinction, \citetalias{1989ApJ...345..245C} gives higher values of $E(B-V)$ by $\sim 0.02$\,mag. This may be related to a lack of UV data at higher extinction, which gives more weight to the optical data where the two extinction laws differ.

The uncertainties on the fitted $E(B-V)$ are between 0.01 and 0.04 mag in most cases, and around 0.1\,mag at higher extinction and/or when fewer data points are available. The results suggest that we can discern extinctions as small as $E(B-V) = 0.02$\,mag under the right conditions. 

Our approach generally finds relatively low extinctions. The highest extinction fitted to a reliable CSPN candidate is for the central star of Hen 2-11, which is found to have  $E(B-V) = 1.47^{+0.04}_{-0.23}$, albeit from a fit with only three SED points. The highest extinction found to a CSPN SED with $\ge5$ data points and naive reduced $\chi^2_{\rm{r}} < 5$ is to the central star of Abell 59, with $E(B-V) = 1.17\substack{+0.07 \\ -0.06}$. The lack of higher extinctions is likely a reflection on the sample: extended PNe are not easily detectable at high extinction and their CSPNe become too faint for \emph{Gaia}.

19 CSPNe give marginally negative extinction results. This provides a useful indicator of the precision of our method. Given the assigned error budget, we find that these 19 points are consistent with random scatter around zero extinction, suggesting that the quoted uncertainties are realistic.

\subsubsection{Comparison to other extinction determinations}
\label{sec:ext}

The CSPNe-derived extinctions can be compared to previously published values for the nebulae. The most complete compilation is by \citetalias{2016MNRAS.455.1459F}, which contains Balmer decrement, optical--radio continuum, CSPNe and dust map extinctions. Some Balmer decrement extinctions are recalculated 
from the literature, and some of the given extinctions are averaging the results of different existing determinations. In our sample of 262 PNe, the \citetalias{2016MNRAS.455.1459F} compilation contains extinctions for 186 sources. Considering only PNe for which good CSPNe candidates were identified and good fits obtained 
leaves 79 sources in common.  \citetalias{2016MNRAS.455.1459F} adopt the extinction law of \citet{1983MNRAS.203..301H} and provide $E(B-V)$ values.  

The extinction curves of \textsc{G-Tomo} are independent of observations of the PNe, so provide a useful comparison  when a distance can be determined. 

\begin{figure}
\includegraphics[width=\columnwidth]{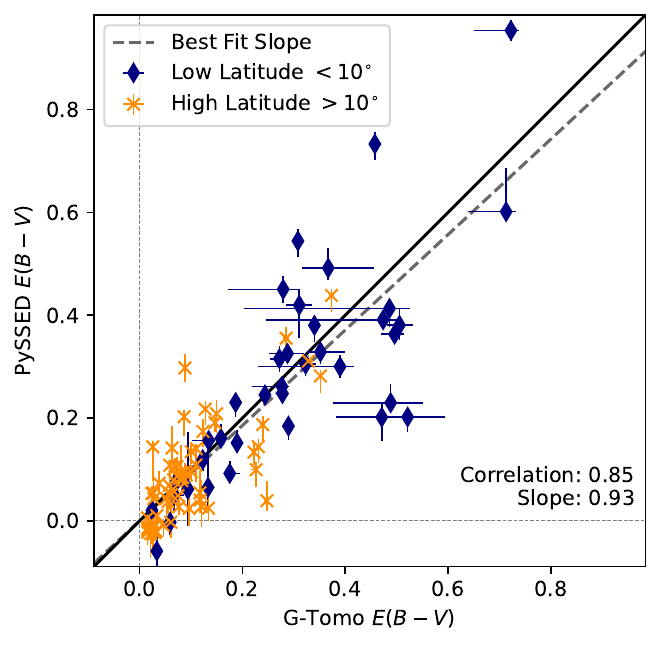}
\caption{Comparison of the $E(B-V)$ results from \textsc{PySSED} with the extinctions derived from \textsc{G-Tomo}. The uncertainties in the  presented \textsc{G-Tomo} $E(B-V)$ are derived from the distance uncertainty to the PN. Only \textsc{PySSED} fits with $\ge5$ SED points and a naive $\chi_r^2<5$} are included.
\label{fig:PySSED_G-Tomo_Comp}
\end{figure}

\begin{figure}
\includegraphics[width=\columnwidth]{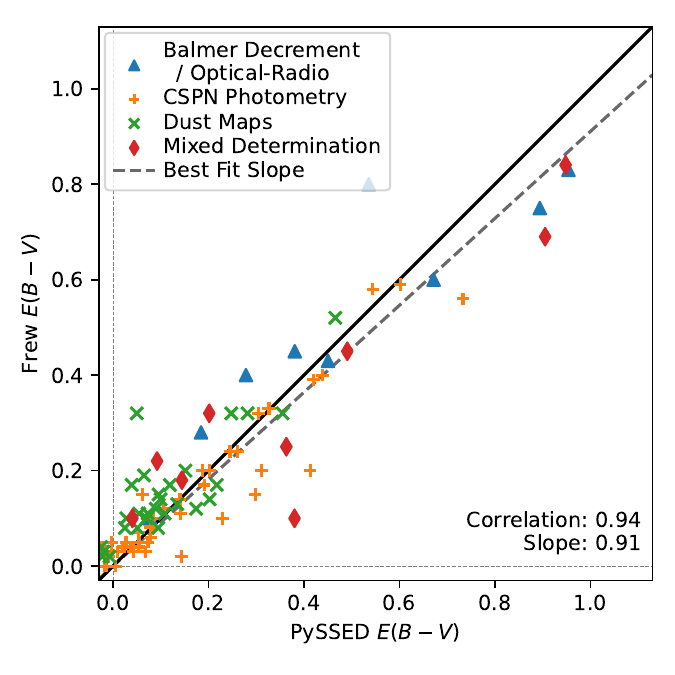}
\caption{Comparison of the $E(B-V)$ results from \textsc{PySSED} with the compilation of extinctions of \citetalias{2016MNRAS.455.1459F}. Extinctions from the Balmer decrement or Optical--Radio comparison are grouped together in the compilation and not distinguished. Mixed determination here refers to cases where an extinction value is averaged from multiple methods. Only \textsc{PySSED} fits with $\ge5$ SED points and a naive $\chi_r^2<5$ are included.}
\label{fig:PySSED_Frew_Comp}
\end{figure}

Comparisons of the results of this paper with \textsc{G-Tomo} and \citetalias{2016MNRAS.455.1459F} are shown in Fig.~\ref{fig:PySSED_G-Tomo_Comp} and Fig.~\ref{fig:PySSED_Frew_Comp}, respectively.

We compare with a number of other sets of extinctions, detailed in Table \ref{tab:ExtCompilations}. \citet{2013MNRAS.431....2F} find Balmer decrement extinctions by comparing their integrated H$\alpha$ fluxes with integrated H$\beta$ fluxes they compile from the literature.  \citet{1992AnAS...94..399C} present two extinction measures: a radio--H$\beta$ extinction found using their 5 GHz radio continuum fluxes and integrated ${\rm H}\beta$ fluxes, and a Balmer decrement extinction found by additionally utilising integrated ${\rm H}\alpha$ fluxes from the literature. The radio--optical extinctions of \citet{1998ApJS..117..361C} are found using their 1.4 GHz radio continuum fluxes and  ${\rm H}\beta$ fluxes from the ESO--Strasbourg catalogue of PNe. We calculate Balmer decrement extinctions from the line ratios presented by the ESO--Strasbourg catalogue \citep{1992secg.book.....A}.

\begin{table*}
\centering
\caption{Sets of literature extinction determinations with which the results are compared. The 'extinctions found' column refers to the number of sources within our sample of 262 extended planetary for which an extinction is given.}
\label{tab:ExtCompilations}
\begin{tabular}{llll}
\toprule
Shorthand name & Method & Extinctions found & Reference  \\
\midrule
\textsc{G-Tomo} & Tomographic extinction & 166 & \cite{2022AnA...661A.147L, 2022yCat..36640174V} \\
Frew Compiled & Mixed & 186 & \cite{2016MNRAS.455.1459F} \\
Frew Balmer & Balmer decrement & 20 & \cite{2013MNRAS.431....2F} \\
Cahn Balmer & Balmer decrement & 55 & \cite{1992AnAS...94..399C} \\
Cahn Radio & Radio--optical & 27 & \cite{1992AnAS...94..399C} \\
Condon Kaplan & Radio--optical & 31 & \cite{1998ApJS..117..361C} \\
ESO/Strasbourg & Balmer decrement & 46 & \cite{1992secg.book.....A} \\
\bottomrule

\end{tabular}
\end{table*}

\begin{figure}
{\includegraphics[width=\columnwidth]{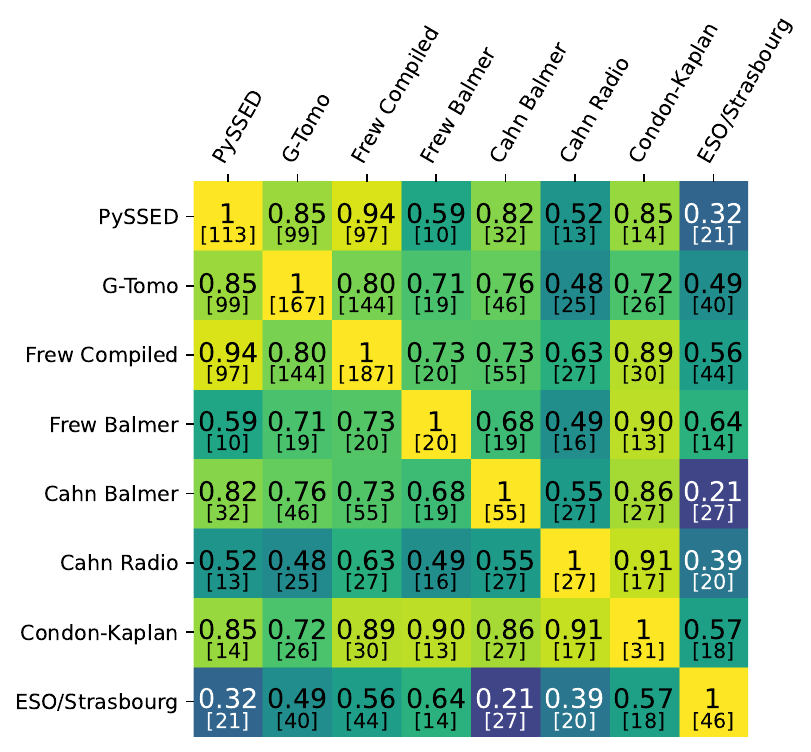}}
\caption{The Pearson correlation coefficient between extinction compilations. The lower number in square brackets is the number of sources the correlation is calculated between. The correlation is only calculated between PNe available in both data sets. The colour represents the difference from $1$, perfect agreement, ranging from yellow to dark blue. The shorthand for the datasets is given in Table \ref{tab:ExtCompilations}.} 
\label{fig:Correlation}
\end{figure}

\begin{figure}
{\includegraphics[width=\columnwidth]{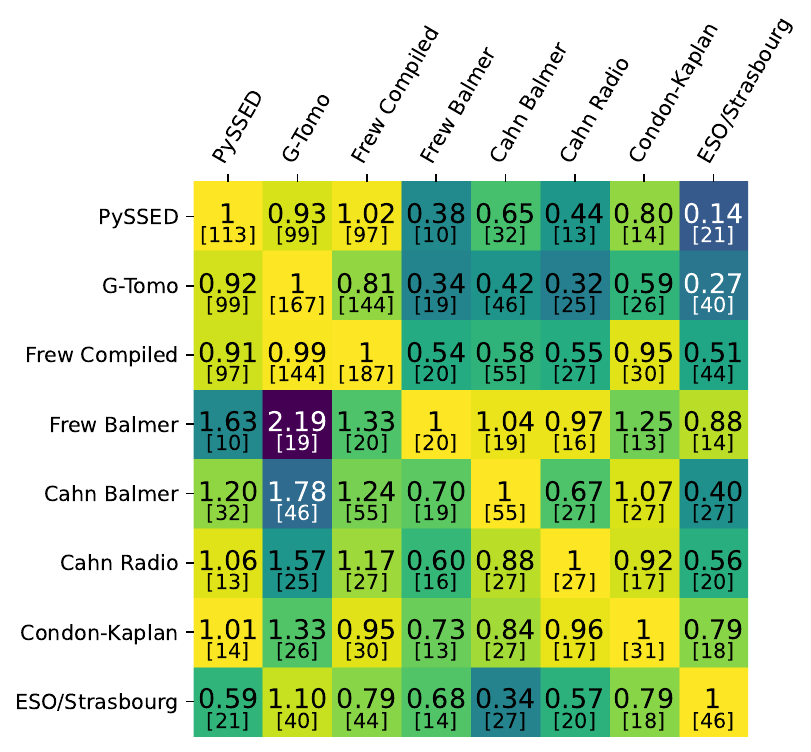}}
\caption{The scale factor $m$ between the extinction determinations, calculated by least-squares minimisation of $C_{\beta, j} = m C_{\beta, i}$. In each case, $i$ is the row and $j$ is the column. The lower number in square brackets is the number of sources the scale is calculated between. The colour represents difference from $1$ (perfect agreement), ranging from yellow to dark blue. The shorthand for the datasets is given in Table \ref{tab:ExtCompilations}.}  
\label{fig:Slope}
\end{figure} 

Figure \ref{fig:Correlation} shows the Pearson correlation coefficients between the different data sets. Since this coefficient does not depend on scaling factor, it is independent of the actual representation used ($C_\alpha$, $C_\beta$, $E(B-V)$). These correlations are calculated for sources with a good quality fit: when including all 162 fits the correlations with \textsc{G-Tomo} and the Frew Compiled extinctions weakens to 0.87.

A good correlation does not ensure a 1:1 correspondence. For all correlations in Fig.~\ref{fig:Correlation} a least-square-minimisation linear fit was done where the intercept was fixed to the origin. The resulting slopes are shown in Fig.~\ref{fig:Slope}. For the main surveys the slopes are indeed close to unity.

The \textsc{PySSED}, \textsc{G-Tomo} and Frew Compiled extinctions show good agreement with each other, with correlations close to one and similar scales. This close correlation across different methodologies suggests that these approaches are successfully measuring extinction. For a limited set of the brightest objects in the sample, the radio extinctions of \citet{1998ApJS..117..361C} also show excellent correlation with other indicators. We note that the number of sources in the correlations is relatively small in all but the correlations between PySSED, \textsc{G-Tomo} and Frew Compiled.

Other extinction indicators appear to be less accurate. The radio extinctions of \citet{1992AnAS...94..399C} are not well correlated with the other results. The extinctions derived from the line ratios in the ESO/Strasbourg catalogue show poor correlations and sometimes indicate significant negative extinctions up to $E(B-V)=-1$\,mag. The Balmer decrement extinctions of \citet{2013MNRAS.431....2F} are reasonably correlated, but show a significant offset in scale when compared to other sources: \citet{2013MNRAS.431....2F} report much higher extinctions for some PNe than found in other surveys. The same offset is seen between \citet{2013MNRAS.431....2F} and the compilation of \citetalias{2016MNRAS.455.1459F}, suggesting that the previous values have been superseded by the later work. 

The correlations can be affected by sensitivity. Fainter nebulae are more difficult to measure and may have less accurate data. We selected nebulae larger than 1 arcmin, which can be difficult for spectroscopy because of low surface brightness, and for integrated flux measurements because of the large required aperture and inclusion of more background emission.  
Extinction fitting to CSPNe avoids these issues, making it the most reliable extinction indicator for extended nebulae of low surface brightness.

\subsection{Luminosities}
\label{sec:luminosities}

Luminosities have been determined where a distance to the CSPN or PN is available. This is performed using the ordinary luminosity determination procedure of \textsc{PySSED}, detailed in \citet{2024RASTI...3...89M}. The parallax-derived \textsc{PySSED} distances are used if the fractional error of the parallax is smaller than 0.5. Otherwise, the \citetalias{2016MNRAS.455.1459F} surface-brightness distances are used if available. If neither is available, we do not calculate a luminosity.

The PySSED luminosities assume $T_{\rm eff}= 10^5$\,K ($L_\text{100kK}$). If a more accurate $T_{\rm eff}$ is available, we use the scaling $L \propto T^3$ to find a temperature corrected luminosity $L_\ast$\footnote{Bolometric luminosity scales as $T^4$, while at constant $L$, the observed Rayleigh--Jeans-tail  flux scales as $T$, hence $L \propto T^3$}. We have performed an updated compilation of literature CSPNe temperature determinations for our A/A+ stars, based on the compilation of \citet{2020A&A...640A..10W}, adding new temperatures where available. Temperatures derived from spectroscopy of the star are most accurate, and prioritised in our compilation. He\,{\sc II} Zanstra temperatures are included, but hydrogen Zanstra values are rejected since our extended nebulae are unlikely to be fully optically thick to H-ionizing radiation. For each star, we have identified and referenced the original source of the temperature determination.
The literature temperature compilation and temperature corrected luminosities $L_\ast$ are presented in Table \ref{tab:scaled_lum_sources}. We were able to calculate temperature corrected luminosities for 49 of the sources.  The temperature uncertainty is taken from the literature where available;  a 20\% uncertainty is assumed if no uncertainty is given. The uncertainty on the temperature corrected luminosity $L_\ast$ is taken to be dominated by the uncertainties on the temperature and distance. The typical temperature derived uncertainty on the uncorrected luminosity $L_{100\rm{kK}}$ of the remaining sources is estimated at $0.5$\,dex based on a typical temperature error of a factor of 1.5, assuming the temperature range for these sources is similar to the range of spectroscopic temperatures.

\begin{figure*}
\includegraphics[width=\textwidth]{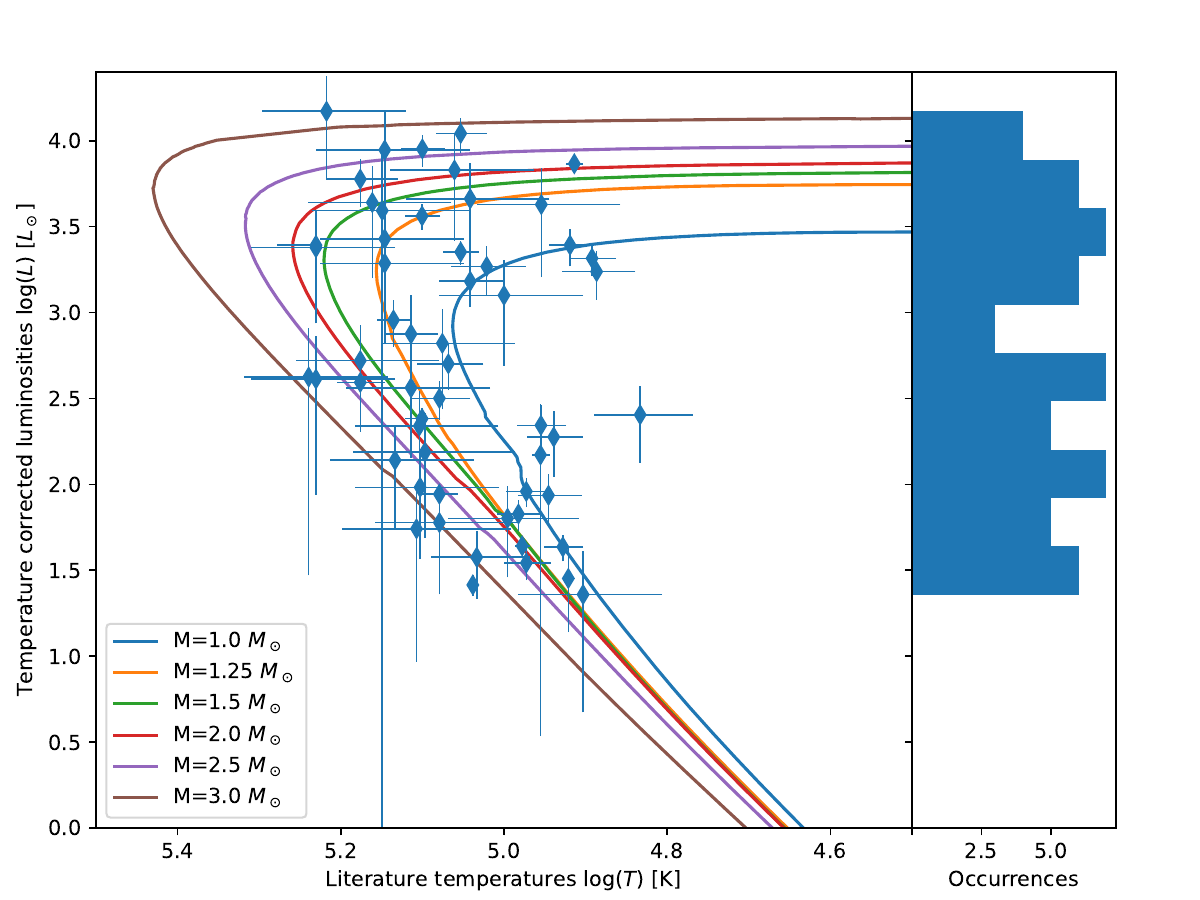}
\caption{An HR diagram made by adjusting  $L_{\text{100kK}}$ to the literature temperatures of Table \ref{tab:scaled_lum_sources} using $L \propto T^3$. The modelled CSPN cooling tracks of \citet{2016A&A...588A..25M} are included, where the tracks are differentiated based on the initial mass of the star. cf., Figure 18 of \citet{2021A&A...656A..51G}, which uses a different approach to present a similar HR diagram based on luminosities calculated by bolometric correction of the $V$ band magnitude. }
\label{fig:LumPlots}
\end{figure*}

\begin{figure}
{\includegraphics[width=\columnwidth]{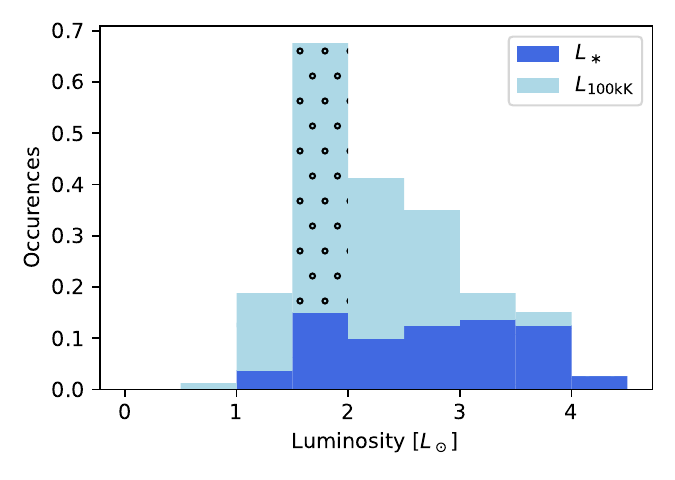}}
\caption{A histogram of the best available luminosity determination for each source. The temperature corrected luminosities $L_\ast$ are used where available, $L_\text{100kK}$ is used where a temperature cannot be determined.}  
\label{fig:LumHist}
\end{figure}

Fig.~\ref{fig:LumPlots} shows the temperature corrected luminosities $L_\ast$ plotted against the literature temperatures in a Hertzsprung-Russell diagram. The results agree well with the theoretical CSPN evolutionary tracks produced by the models of \citet{2016A&A...588A..25M}. This figure can be compared with Fig.~18 of \citet{2021A&A...656A..51G} which shows a similar HR diagram based on luminosities calculated by bolometric correction of the $V$
band magnitude. Their sample includes many more stars on the horizontal track as they do not limit their study to extended nebulae. Figure \ref{fig:LumHist} shows a histogram of the best-fit luminosities for all of the objects within the sample (i.e., the temperature corrected used where available, the $L_\text{100kK}$ used otherwise). 

\begin{table*}
\centering
\caption{Objects for which a literature temperature is available, therefore a temperature corrected luminosity can be obtained. 
The sample is based on the compilation of \citet{2020A&A...640A..10W}, with updated literature temperatures where available.}
\begin{tabular}{llllllc}
\toprule
PN G & PN Name & $\rm L_{\text{100kK}}$ [$\rm L_\odot$] & Literature Temperature$^{\rm a}$ [$\rm K$] & Source & Distance &  Temperature Corrected Luminosity [$\rm L_\odot$] \\
\midrule
002.7-52.4 & IC 5148/50 & 170 & $130\,000$ & BK2018 & $1170\pm 70$ & $360 \pm 220$ \\
017.3-21.9 & Abell 65 & 1100 & $110\,000 \pm 10\,000$ & HF2015 & $1420\pm 70$ & $1520 \pm 440$ \\
025.4-04.7 & IC 1295 & 300 & $90\,100 \pm 6200$ & N1999 & $1600\pm 200$ & $221 \pm 71$ \\
034.1-10.5 & HaWe 13 & 800 & $68\,100 \pm 9400$ & N1999 & $1800\pm 200$ & $254 \pm 121$ \\
035.9-01.1 & Sh 2-71 & 80 & $173\,525$ & B2008 & $1500 \pm 500 \dagger$ & $420 \pm 390$ \\
036.0+17.6 & Abell 43 & 3400 & $110\,000$ & DW1997 & $2090\pm 110$ & $4590 \pm 2800$ \\
036.1-57.1 & NGC 7293 & 50 & $120\,000 \pm 6000$ & TH2005 & $199\pm 2$ & $88 \pm 13$ \\
041.8-02.9 & NGC 6781 & 80 & $96\,000 \pm 6000$ & K1983 & $494\pm 19$ & $67 \pm 14$ \\
046.8+03.8 & Sh 2-78 & 30 & $120\,000$ & D1999 & $700\pm 50$ & $60 \pm 37$ \\
047.0+42.4 & Abell 39 & 300 & $117\,000 \pm 11\,000$ & N1999 & $1160\pm 50$ & $503 \pm 149$ \\
055.4+16.0 & Abell 46 & 4300 & $83\,000 \pm 5000$ & SP2008 & $2500\pm 200$ & $2470 \pm 610$ \\
059.7-18.7 & Abell 72 & 500 & $170\,000$ & BW2023 & $1840\pm 200$ & $2390 \pm 1520$ \\
060.8-03.6 & NGC 6853 & 120 & $126\,000 \pm 6300$ & TH2005 & $387\pm 7$ & $241 \pm 37$ \\
063.1+13.9 & NGC 6720 & 110 & $127\,000$ & P1996 & $780\pm 30$ & $219 \pm 132$ \\
066.7-28.2 & NGC 7094 & 4500 & $125\,900 \pm 7700$ & DW1997 & $1610\pm 80$ & $8960 \pm 1890$ \\
072.7-17.1 & Abell 74 & 30 & $108\,000 \pm 15\,000$ & N1999 & $670\pm 30$ & $38 \pm 16$ \\
076.3+14.1 & Pa 5 & 1400 & $145\,000$ & DL2015 & $2300\pm 200$ & $4370 \pm 2780$ \\
077.6+14.7 & Abell 61 & 130 & $88\,200 \pm 7900$ & N1999 & $1630\pm 160$ & $86 \pm 29$ \\
080.3-10.4 & MWP 1 & 500 & $170\,000\substack{+20\,000\\ -1000}$ & RW1997 & $499\pm 9$ & $2470 \substack{+870 \\ -100}$ \\
081.2-14.9 & Abell 78 & 8000 & $113\,000 \pm 8000$ & HB2004b & $1640\pm 80$ & $11\,030 \pm 2580$ \\
085.3+52.3 & Jacoby 1 & 120 & $150\,000 \pm 10\,000$ & AW2012 & $780\pm 30$ & $392 \pm 84$ \\
094.0+27.4 & K 1-16 & 1000 & $140\,000$ & KW1998 & $1740\pm 90$ & $2680 \pm 1630$ \\
104.2-29.6 & Jn 1 & 160 & $150\,000$ & RW1995 & $980\pm 60$ & $530 \pm 320$ \\
118.8-74.7 & NGC 246 & 1800 & $150\,000 \pm 15\,000$ & RW1997 & $538\pm 19$ & $5970 \pm 1840$ \\
136.1+04.9 & Abell 6 & 300 & $86\,900 \pm 6800$ & RB2024 & $1300\pm 200$ & $189 \pm 78$ \\
144.8+65.8 & LTNF 1 & 1600 & $105\,000 \pm 11\,000$ & LT1995 & $1350\pm 40$ & $1850 \pm 590$ \\
148.4+57.0 & NGC 3587 & 110 & $93\,900 \pm 5600$ & N1999 & $820\pm 30$ & $91 \pm 17$ \\
149.4-09.2 & HaWe 4 & 80 & $125\,000 \pm 28\,000$ & N1999 & $910\pm 60$ & $154 \pm 105$ \\
153.7+22.8 & Abell 16 & 50 & $83\,400 \pm 1500$ & RB2024 & $1500\pm 400$ & $28 \pm 15$ \\
158.6+00.7 & Sh 2-216 & 50 & $95\,000 \pm 2000$ & ZR2009 & $127\pm 1$ & $44 \pm 3$ \\
158.9+17.8 & PuWe 1 & 40 & $93\,900 \pm 6200$ & N1999 & $394\pm 9$ & $35 \pm 7$ \\
164.8+31.1 & JnEr 1 & 50 & $126\,712$ & B2008 & $980\pm 80$ & $96 \pm 60$ \\
205.1+14.2 & Abell 21 & 50 & $136\,000$ & F2008 & $580\pm 20$ & $138 \pm 84$ \\
208.5+33.2 & Abell 30 & 4400 & $115\,000$ & LK1993 & $2080\pm 130$ & $6750 \pm 4140$ \\
214.9+07.8 & Abell 20 & 400 & $119\,000 \pm 22\,000$ & RK1999 & $1850\pm 180$ & $660 \pm 390$ \\
215.5-30.8 & Abell 7 & 70 & $99\,000 \pm 18\,000$ & N1999 & $512\pm 11$ & $63 \pm 35$ \\
217.1+14.7 & Abell 24 & 20 & $109\,200 \pm 1000$ & RB2024 & $750\pm 50$ & $26 \pm 4$ \\
219.1+31.2 & Abell 31 & 70 & $84\,700 \pm 4700$ & N1999 & $533\pm 15$ & $43 \pm 8$ \\
255.3-59.6 & Lo 1 & 180 & $120\,000 \pm 10\,000$ & HB2004a & $800\pm 20$ & $317 \pm 81$ \\
258.0-15.7 & Lo 3 & 700 & $140\,000$ & RW1995 & $2100\pm 300$ & $1930 \pm 1270$ \\
259.1+00.9 & Hen 2-11 & 3200 & $140\,000 \pm 30\,000$ & JB2014 & $1600\pm 170$ & $8800 \pm 6000$ \\
283.6+25.3 & K 1-22 & 1400 & $141\,000 \pm 31\,000$ & RK1999 & $1500\pm 700$ & $3920 \substack{+4310 \\ -3920}$ \\
283.9+09.7 & DS 1 & 3800 & $77\,000 \pm 8000$ & D1985 & $789\pm 19$ & $1730 \pm 550$ \\
293.6+10.9 & BlDz 1 & 30 & $128\,000 \pm 30\,000$ & RK1999 & $1400\pm 300$ & $55 \pm 46$ \\
294.1+43.6 & NGC 4361 & 1800 & $126\,000 \pm 6300$ & TH2005 & $1000\pm 40$ & $3650 \pm 620$ \\
307.2-03.4 & NGC 5189 & 3300 & $165\,000$ & KB2014 & $1410\pm 30$ & $14\,800 \pm 8900$ \\
310.3+24.7 & Lo 8 & 6000 & $90\,000$ & HM1990 & $1460\pm 120$ & $4250 \pm 2640$ \\
318.4+41.4 & Abell 36 & 1600 & $113\,000 \pm 5650$ & TH2005 & $408\pm 9$ & $2250 \pm 350$ \\
329.0+01.9 & Sp 1 & 13\,000 & $82\,000 \pm 2000$ & SK1989 & $1320\pm 20$ & $7350 \pm 600$ \\
332.5-16.9 & HaTr 7 & 1300 & $100\,000$ & SW1997 & $1610\pm 90$ & $1260 \pm 770$ \\
335.5+12.4 & DS 2 & 4400 & $78\,000 \pm 5000$ & MK1987 & $800\pm 40$ & $2070 \pm 440$ \\
276.2-06.6 & PHR J0907-5722 & 40 & $80\,000$ & AW2019 & $1700\pm 400$ & $23 \pm 18$ \\
066.5-14.8 & Kn 45 & 200 & $90\,200 \pm 2300$ & RB2024 & $2800\pm 1400$ & $148 \pm 145$ \\
079.8-10.2 & Alves 1 & 80 & $170\,000$ & RB2024 & $1900\pm 500$ & $410 \pm 320$ \\
098.3-04.9 & Pa 41 & 400 & $136\,600 \pm 6600$ & RB2024 & $2200\pm 300$ & $900 \pm 270$ \\
134.3-43.2 & Pre 8 & 300 & $130\,000 \pm 9600$ & RB2024 & $3500\pm 1100$ & $750 \pm 520$ \\
\bottomrule
\end{tabular}
    \newline $^{\rm a}$Uncertainties are given from the literature where available, otherwise a 20\% uncertainty is adopted. $\dagger$ The \citetalias{2016MNRAS.455.1459F} distance has been used for the nebula.
    \newline BK2018: \citet{2018A&A...620A..84B}; HF2015: \citet{2015AJ....150...30H}; N1999: \citet{1999A&A...350..101N}; B2008: \citet{2008ApJ...674..954B}; TH2005: \citet{2005ASPC..334..325T}; DW1997: \citet{1997IAUS..180..103D}; K1983: \citet{1983ApJ...271..188K}; D1999: \citet{1999RvMA...12..255D}; SP2008: \citet{2008AstL...34..423S}; BW2023: \citet{2023MNRAS.521..668B}; P1996: \citet{1996A&A...307..561P}; DL2015: \citet{2015MNRAS.448.3587D}; RW1997: \citet{1997fbs..conf..217R}; HB2004b: \citet{2004ApJ...609..378H}; AW2012: \citet{2012A&A...546A...1A};
    KW1998: \citet{1998ApJ...502..858K}; RW1995: \citet{1995LNP...443..186R};
    RB2024: \citet{2024A&A...690A.366R}; LT1995: \citet{1995ApJ...441..424L}; ZR2009: \citet{2009Ap&SS.320..257Z}; F2008: \citet{2008PhDT.......109F}; LK1993: \citet{1993AcA....43..329L};  RK1999: \citet{1999A&A...347..169R}; HB2004a: \citet{2004PASP..116..391H}; JB2014: \citet{2014A&A...562A..89J}; D1985: \citet{1985ApJ...294L.107D}; KB2014: \citet{2014MNRAS.442.1379K}; HM1990: \citet{1990Ap&SS.169..183H}; SK1989: \citet{1989ApJS...69..495S}; SW1997: \citet{1997A&A...328..598S}; AW2019: \citet{2019ApJ...882..171A}.
\label{tab:scaled_lum_sources}
\end{table*}

Roughly half of the CSPNe for which temperature corrected luminosities could be obtained are located on the cooling track, with luminosities between 20 and  500\,L$_\odot$. This is expected, since the PNe were selected based on a large angular size with likely higher nebular ages.  The remainder are consistent with the horizontal part of the post-AGB tracks. Almost all CSPNe have $T>80$\,kK and most have $L<10^3$\,$L_\odot$, consistent with higher ages. The stars for which temperatures are available may be biased towards the brighter stars, therefore including a larger fraction of stars on the higher-luminosity horizontal track than found in the complete sample shown in Fig.~\ref{fig:LumHist}.

A few objects are on the horizontal tracks at high luminosities, which would imply short evolutionary timescales, inconsistent with the large size. Of these, the luminosity of NGC 5189 has a large uncertainty. Abell 78 is a born-again PN which has evolved from the cooling track \citep{2015ApJ...799...67T} and is therefore much more evolved than indicated by its location on the HR diagram. Some objects have very uncertain determinations resulting from their distance uncertainties, e.g. K 1-22, which we note has some issues with its \textit{Gaia} identifier.

\subsection{Individual objects}

The results from some stars warrant further discussion. Abell 46 is an object with a range of literature temperature determinations. The CSPNe of HaWe 13 and IC 1295 are significantly subluminous in Fig.~\ref{fig:LumPlots}. The extinction results of NGC 6781 and Hen 2-11 are in notable disagreement with the extinctions from \textsc{G-Tomo}.

\subsubsection{Abell 46}
\label{Sec:A46}
\begin{figure}
\includegraphics[width=\columnwidth]{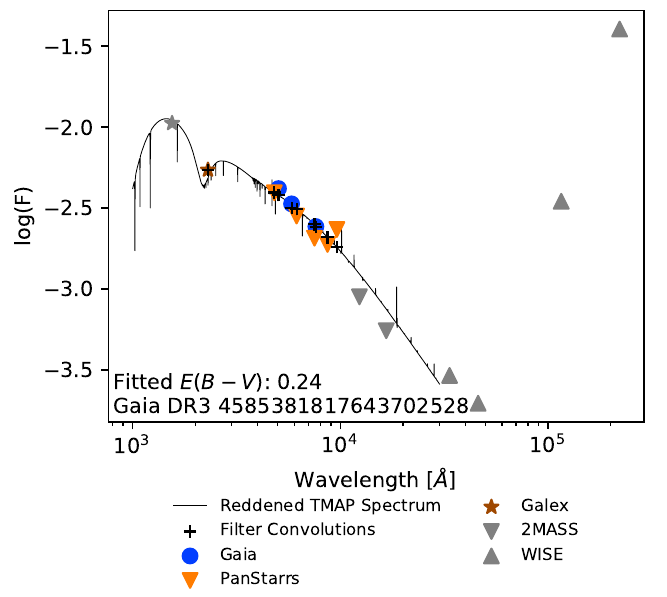}
\caption{SED of the CSPN of Abell 46.}
\label{fig:Abell46_SSED}
\end{figure} 

The central star of Abell 46, designated V477 Lyr, has been identified as a candidate post-RGB PN \citep{2017AJ....153...24H}, based on the location of its CSPN in the HR diagram. The stellar system is an eclipsing close binary, consisting of a CSPN alongside a late K- or early M-type companion with an orbital period of 0.47 days. The literature is not in agreement regarding the CSPN temperature. \citet{2008MNRAS.391..802A} derive respective photometric temperatures of $49\,500 \pm 4500$\,K and $3874 \pm 350$\,K for the CSPN and companion respectively. 
This leaves the star significantly subluminous on the HR diagram. However, a much higher spectroscopic temperature for the CSPN of $83\pm5$\,kK is derived by \citet{2008AstL...34..423S}. 

The cool companion has a hot spot on its illuminated side,  with a temperature as high as 20\,kK \citep{2008AstL...34..423S}, causing significant orbital variability. The hot spot dominates the emission from the cool star even in the optical. Our SED (Fig.~\ref{fig:Abell46_SSED}) does not show clear evidence for the cool star, while the hot spot and the CSPN cannot easily be separated. 
We do not separate the sources, but rather assume that for the data $<10\,000$\,\AA \, the CSPN dominates the luminosity, allowing us to do a single-star SED fit. This fit ignores the hot spot on the companion, and should thus be considered indicative rather than an accurate measurement of extinction or luminosity.

For the higher temperature of $83$\,kK, we find $E(B-V)=0.24$ and a temperature corrected luminosity $\rm L_\ast=2700\,\rm L_\odot$. 
This places the CSPN just below the evolutionary tracks of \citet{2016A&A...588A..25M}. For the lower temperature of $T=49\,500$\,K \citep{2008MNRAS.391..802A} we repeated the fit using the TMAP model at $T=50$\,kK, as the temperature  falls outside the range where  our fitting procedure is robust. This fit gives $\rm L_\ast = 490\,\rm L_\odot$ and $E(B-V)=0.19$. This would place the star well below the post-AGB tracks. 

The classification of Abell\,46 as a post-RGB system is therefore not certain: at the higher stellar temperature the PN is compatible with post-AGB evolution.  The surface-brightness distance of \citetalias{2016MNRAS.455.1459F}  agrees with the \emph{Gaia} distance which would be unexpected for post-RGB systems. 

\subsubsection{HaWe 13}

The most sub-luminous object in our sample is HaWe 13. It has also been noted as a post-RGB PN candidate \citep[e.g.,][]{2017AJ....153...24H}. Our luminosity determination for the CSPN is based on a temperature of $70$\,kK from \citet{1999A&A...350..101N}, wherein the nebula is referred to as HDW 11. 
The nebula is poorly studied.  \citet{1999A&A...350..101N} finds different temperatures from different stellar lines. We recommend deeper spectroscopy of this star to confirm its location on the HR diagram.

\subsubsection{NGC 6781}

In the case of NGC 6781, the results of the various extinction methods appear to be in disagreement. The \textsc{PySSED} extinction is $C_\beta=0.79^{+0.03}_{-0.05}$, which agrees with the extinction $C_\beta=0.84$ of \citetalias{2016MNRAS.455.1459F} based on CSPN photometry. \textsc{G-Tomo} gives a significantly lower value of $C_\beta=0.45 ^{+0.01}_{-0.02}$.

The observed Balmer decrements of \citet{2001A&A...374..280M} indicate that there is differing extinction across the nebula, varying from $C_\beta = 0.45$ in the south to $C_\beta = 0.59$ in the north (conversion performed using \citetalias{1999PASP..111...63F} with the values in Table \ref{tab:ExtLaws}). The low value agrees very well with the interstellar extinction from \textsc{G-Tomo}. The high value is lower than that found by \textsc{PySSED} and in \citetalias{2016MNRAS.455.1459F}.  The central star thus has high extinction compared to most of the PN\footnote{\citet{1987ApJS...65..405A} report $C_\beta=1.0$ based on the Balmer decrement, consistent with the stellar extinction but it is not clear where in the nebula this was measured.}.

The \emph{GALEX} image in the HASH database shows only the southern part of the nebula, in addition to the star. If the rest of the nebula is not detected because of extinction, then this extinction screen indeed covers the location of the central star. The variation across the face of nebula indicates that the extra extinction is near the PN and is possibly associated with it. The physical size of NGC 6781 is 0.26\,pc across its minor axis, which  makes it the second most compact object in the sample for which the physical size can be calculated.

\subsubsection{Hen 2-11}

Hen 2-11 is a bipolar PN with a close-binary post-common envelope system at the centre \citep{2014A&A...562A..89J}. The central star has a \emph{Gaia} parallax-derived distance of $1600 \pm 170$ pc, which is in disagreement with the surface brightness distance of $660 \pm 130$ pc from \citetalias{2016MNRAS.455.1459F}.

The \textsc{G-Tomo} dust maps return a $C_\beta = 0.89 \pm 0.10$ at the \emph{Gaia} parallax-derived distance. The SED is sparse, containing only the three photometric points from \emph{Gaia}, however the resulting $C_\beta=2.13 \substack{+0.34 \\ -0.05}$ is comparable to the other extinction results derived for the nebula: \citetalias{2016MNRAS.455.1459F} obtain $C_\beta = 2.29$ from a combination of the Balmer decrement and central-star photometry, \citet{1992AnAS...94..399C} find $C_\beta = 2.24$ from the Balmer decrement of the nebula, while \citet{1998ApJS..117..361C} derive $C_\beta = 2.1$ from a comparison of radio and ${\rm H}\beta$ fluxes. Additionally, the H$\alpha$/H$\beta$ ratio of \citet{2014AA...562A..89J} gives a $C_\beta = 2.41$.

The higher extinction found for the PN compared to the interstellar dust maps indicates an additional dust cloud. This would have to cover the nebula relatively evenly, as the extinctions derived from integrated fluxes are in broad agreement with the extinction of the central star. The most likely explanation given the low Galactic latitude of the object is interstellar dust. The methodology of the dust maps limits their ability to detect compact regions of high extinction because stars beyond such extinction regions become undetected.

\subsection{Binaries with composite SEDs}
\label{sec:composite_seds}

Some of the objects in the sample were found to have strong UV emission that suggests a CSPN, but where the optical and IR photometry cannot be well fit with a reddened CSPN spectrum and suggests a main-sequence (MS) star. We have designated a class of seven objects within the sample that exhibit this composite pattern (Table \ref{tab:unresolved_binary_results}). The SEDs of these objects can be better fit by assuming that they are binary systems composed of a CSPN and a main-sequence companion. Four of these, We 3-1, PTB 32, Pa 161 and StDr 14, are new  binary candidates.

We do not fit extinctions for these objects as the CSPN photometry cannot be isolated. Instead, the extinction is adopted from \textsc{G-Tomo}. For PTB 32, no \emph{Gaia} distance is available and the statistical distance of \citet{2016MNRAS.455.1459F} is used instead. For NGC 1514, the extinction found by \citet{2010AJ....140.1882R} is adopted.

For each object, depending on the relative luminosities, we can assume particular portions of the SED to be completely dominated by one of the binary components. This component can then be fitted, then subtracted from the rest of the SED. The other component of the binary can then be fitted to this subtracted SED. 

We fit the CSPN component to the UV photometry using the $T=100$ kK, $\log(g)=6.0$ TMAP model. The MS component is fitted to the optical and IR photometry using the standard procedure of \textsc{PySSED} \citep{2024RASTI...3...89M}, i.e., fitting temperature and surface gravity to \textsc{BT-Settl} models. Greater detail and plots of the SEDs for the newly identified binaries are given in Appendix \ref{app:binary_SEDs}.

Table \ref{tab:unresolved_binary_results} gives the parameters which are fit using the SED, namely the temperature and luminosity of the MS companion and the luminosity of the CSPN assuming a temperature of 100\,kK.

\begin{table*}
\caption{The results of the binary fitting procedure for the sources which we identify as unresolved binaries. 
}
\centering
\begin{tabular}{llllllll}
\toprule
PN G & PN Name & \emph{Gaia} DR3 ID & $E(B-V) $& CSPN $L_{\text{100kK}}$ & MS Temperature & MS Luminosity  & Distance \\
 & & & (\textsc{G-Tomo}) & [L$_\odot$] &  [K] & [L$_\odot$] & [pc]\\
\midrule
044.3+10.4 & We 3-1 & 4509807233807831936 & 0.32 & 1300 & 6400 & 18 & $3700 \pm 1000$ \\
011.3$-$09.1 & PTB 32 & 4078149826461265920 & 0.30 & 180 & 6000 & 0.36 & $2600\pm800$$^a$\\
026.1$-$17.6 & Pa 161 & 4183333506776770688 & 0.16 & 130 & 5500 & 4.9 & $982 \pm 14$ \\ 
115.9$-$13.5 & StDr 141 & 393535419406444416 & 0.09 & 110 & 6400 & 0.67 & $2300\pm400$ \\
107.0+21.3 & K 1-6 & 2288467186442571008 & 0.03 & 7.8 & 4500 & 1.0 & $258\pm2$ \\
228.2$-$22.1 & LoTr 1 & 2917223705359238016 & 0.03 & 120 & 4600 & 33 & $1790 \pm 50 $ \\
\midrule
165.5-15.2 & NGC 1514$^b$ &  168937010969340160 & 0.52 & 68  & 9200 & 120 & $446\pm4$ \\
\bottomrule
\label{tab:unresolved_binary_results}
\end{tabular}
\newline $^a$The \emph{Gaia} parallax distance for the CSPN of PTB 32 does not meet our reliability criteria that the fractional parallax error is smaller than 0.5, so the surface brightness distance of \citetalias{2016MNRAS.455.1459F} is used instead. $^b$NGC 1514 has been fit slightly differently, as the companion dominates at all wavelengths where we obtain photometry. The adopted $E(B-V)=0.52$ is derived from the $A_V=1.6$ determination by \citet{2010AJ....140.1882R, 2025AJ....169..236R}, see Sec.~\ref{sec:ngc1514}.
\end{table*}

\subsubsection{K 1-6}

We find a possibility that this nebula has a spurious central star identification, and is therefore not a binary. The nebula has a lopsided appearance, taken as indication of interaction with the interstellar medium by  \citet{2011PASA...28...83F}. The assumed central star is offset from the centre in the direction of the bright part of the nebula, as expected in such a case, and has a proper-motion vector in this direction. It is a wide proper-motion double consisting of an 11th-mag bright star with a 16th-mag companion 5 arcsec away.  This fainter companion is itself a visual double with 1 arcsec separation, but the second component has a different parallax and proper motion and is not related. There is disagreement about the identification of the CSPN: the fainter of the proper-motion pair is taken as the CSPN by both \citetalias{2021A&A...656A..51G} and \citetalias{2021A&A...656A.110C}, but the brighter one is identified as the CSPN by \citet{2011PASA...28...83F}.  The brighter star is a 21.3-day photometric variable \citep{2025ApJ...980..227C} with a K1--3 V spectral type \citet{2021MNRAS.506.4151M}. The \emph{GALEX} source has coordinates close to the brighter star, but the UV emission covers both stars.

There are arguments against the proper-motion pair being associated with the PN.  The pair is nearby (258 pc) whilst the statistical distance of the nebula is much larger, 1500$\pm$300 pc \citepalias{2016MNRAS.455.1459F}.  At the closer distance, the interstellar medium stand-off distance, the point where the the gas pressure of the interstellar medium and wind pressure of the CSPN are equal, is less than 0.1 pc whereas for PNe this will normally be of order 1\,pc  \citep{2007MNRAS.382.1233W}. This would indicate a much lower mass-loss rate of the progenitor star than found in AGB stars. The luminosity derived from the \emph{GALEX} photometry is 7.8\,L$_\odot$, which is lower than expected for a CSPN on the cooling track: cooling ages become very long at these luminosities, longer than the life time of a typical PN. In contrast, the proper motion of the pair places it at the centre of the nebula only a few thousand years ago.

This suggests that the true CSPN may not be related to the proper-motion stars and there may be a chance superposition. Using the statistical distance and deriving a luminosity only from the  \emph{GALEX} data results in $L\approx 10^3$\,L$_\odot$, which places the star on the cooling track. Alternatively, the object may not be a true PN but have a different origin.

\subsubsection{NGC 1514} 
\label{sec:ngc1514}

This is a known close binary, where the photometry of the central system is strongly dominated by the companion, which is found to be a 9700\,K star with an extinction of $A_V=1.6$ by \citet{2010AJ....140.1882R, 2025AJ....169..236R}. 

Because the photometry is dominated by the companion at all wavelengths we probe, a slightly different fitting procedure is used. We assume the optical and IR are entirely dominated by the companion and fit the MS star first. Using the extinction of \citet{2010AJ....140.1882R, 2025AJ....169..236R}, \textsc{PySSED} fits a temperature of $T=9300$\,K and $L=120$\,L$_\odot$. This fits the optical photometry well but leaves some \emph{GALEX} UV excess, which is fitted to find the luminosity of the CSPN. 

For the CSPN we find $L=68$\,L$_\odot$, and for the companion $L=130$\,L$_\odot$. These represent the first absolute luminosities determined for the system. 

\section{Discussion}

\subsection{CSPNe candidature}

We find a total of 162 objects for which we are confident about the central-star classification, out of a total sample of 262 PNe. 

PNe where we find no evidence of a hot CSPN are always excluded from category A. An example is the PN Abell 5, where we find a central star which requires a high extinction if it is a hot star, while the Balmer decrement and \textsc{G-Tomo} both indicate a much lower extinction. The detected star is therefore likely cooler, and although it could be a companion, it is not the CSPN itself.

Comparing our results with \citetalias{2021A&A...656A..51G} and \citetalias{2021A&A...656A.110C}, we require, rather than prefer, evidence for a hot star in the photometry. 12 of the objects graded A in \citetalias{2021A&A...656A..51G} are graded B or lower by this work.  Three cases are listed as F and we have no doubt about their rejection. 20 objects which we rate as A or A+ are classed as B by \citetalias{2021A&A...656A..51G}. The additional confidence comes from the evidence for a hot star.   Overall, this indicates fair agreement.
\citetalias{2021A&A...656A.110C} give their CSPN reliability as a number. Of our 160 A category stars in common, 139 are given reliability levels $>0.99$ and 148 are given $>0.95$.

The current sample contains eight CSPNe that are not in the previous lists, of which two are in our A list and an additional one is found to be a binary in Sec.~\ref{sec:composite_seds}. 

\subsection{Multiplicity}

We find seven objects with composite SEDs, which do not have fitted extinctions because of the co-presence of a cool star in the photometry.  (For one of these, K 1-6, there is some doubt whether the binary is related to the PN.) 

There are more CSPN which are known to be binaries through other methods. Combining the catalogue of binary CSPNe maintained by David Jones\footnote{\url{https://www.drdjones.net/bcspn/}} with our CSPNe yields 44 stars in common \citep{2017NatAs...1E.117J, 2019ibfe.book.....B}. Of these, 35 stars are in our A category. Three of these are also in our list of seven composite SEDs. Four stars that show composite SEDs are not in the Jones catalogue: We\,3-1, PTB\,32, Pa\,161 and StDr\,141.  Abell\,46 (Sect. \ref{Sec:A46}) is in the binary catalogue but is not in  Table \ref{tab:unresolved_binary_results} because the cool MS is outshone by the CSPN and hot spot, and is not discernible in the SED. One object from the binary catalogue, Abell\,41, is a visual binary, but the separation is small enough that our photometry would have included both stars. In total, this gives 38 binaries in a full category-A sample of 162 stars. The known binary fraction is 23\%.

The initial sample selection is unbiassed regarding binarity, as it is based only on PN angular size. A bias may come from stars where  \emph{Gaia} detected the companion, but where the CSPN would have been too faint for it. That is mainly an issue at higher extinction. We will also have missed stars where the emission is dominated by a cool companion at all wavelengths, as these are not classified as a category A CSPN  here, even if at the precise centre of the nebula. This is possible for PNe for which we do not have UV data as they lie outside the region surveyed by \emph{GALEX}.

Considering stars at latitudes more than 10 degrees from the Galactic plane, where extinction is not a limiting factor, there are six binaries identified in Table \ref{tab:unresolved_binary_results}, and 22 stars in the binary catalogue, of which 21 are category A. Three stars are in both lists, giving a total of 24 binaries. The total number of category A stars at these latitudes is 83.
This gives a known binary fraction of 29\% outside the Galactic plane. This is slightly higher than the fraction in the full sample. There is a possible detection bias in Jones' binary catalogue, with 16 of the 22 stars being north of the Galactic plane. For $b>+10$ degrees, where we find 47 A-category CSPN and 17 binaries, the binary fraction is 36\%.

We conclude that the known binary fraction in our sample is between 23\%\ and 36\%. The majority of these are stars where the companion does not show up in the SED. PNe where the companion is brighter than than CSPN at optical wavelengths amount to only 5\%. These are mainly the more massive main-sequence stars. 

The fraction of main-sequence stars with binary companions in the solar neighbourhood ranges from 30\%\ for 1\,M$_\odot$ stars to 80\%\ at 4\,M$_\odot$ \citep{Offner2023}. This includes some low-mass, long-period binaries, which may be missed in the PN surveys. The fraction of binarity in our sample is not inconsistent with that of main-sequence stars. 

For two of the binary CSPN in our sample, a  post-RGB track has been proposed as an evolutionary route. For Abell 46, we find that the stellar  temperature determination of \citet{2008AstL...34..423S} can fit a post-AGB track. For HaWe 13, no study has been able to establish the binarity of the CSPN \citep{2023hsa..conf..216J} and the stellar temperature of \citet{1999A&A...350..101N} still needs confirmation. Considering the other post-RGB candidates of \citet{2023hsa..conf..216J}, only HaTr 4 is in our sample, and it does not stand out as having post-RGB properties in Fig.~\ref{fig:LumPlots}. All other objects in our sample are consistent with post-AGB evolution. 

\subsection{Extinction from different methods}

The various methods to determine extinction to PNe measure different aspects. This paper measures the colour changes to broadband photometry over the entire SED, thus traces extinction at a broad range of wavelengths. \textsc{G-Tomo} is based on the interstellar component of the extinction, averaged over a larger area and based on data from the \emph{Gaia} filters which cover the optical broadband spectrum. The Balmer decrement measures the colour changes between the monochromatic Balmer lines only (here H$\alpha$ and H$\beta$), whilst the radio--optical extinction measures the absolute value of the extinction at H$\alpha$ or H$\beta$. The methods measure different aspects of the wavelength-dependent extinction curve, $A_V$ and $R_V$.

The Frew Compiled extinctions \citepalias{2016MNRAS.455.1459F} constitute the most comprehensive set of extinctions. Fig.~\ref{fig:PySSED_Frew_Comp} indicates that the agreement with our results holds across the different methods utilised by the compilation. However, the data set is heterogeneous and is not well documented. Balmer-decrement extinctions are not distinguished from radio extinctions, and the data used for extinction calculations are not indicated. The \textsc{G-Tomo} results depend on distance and are integrated over a wider star field around the nebula, while all other measurements are distance-independent and relate solely to the PN.

The correlation with \textsc{G-Tomo} extinctions are good, both in correlation coefficient and in scale factor. Fig.~\ref{fig:PySSED_G-Tomo_Comp} shows the comparison in detail, where only fits with $\ge5$ SED points and a naive $\chi_r^2<5$ are included. PNe at low and high Galactic latitudes are separated. 
At low extinctions, we find good agreement between \textsc{PySSED} and \textsc{G-Tomo} within $\Delta E(B-V)<0.05$, with some outliers. At higher extinctions, uncertainties increase. There are some outliers where PySSED and \textsc{G-Tomo} disagree by more than 0.2 in $E(B-V)$, but the overall scatter is $<0.1$\,mag. 

At low latitudes $b<10^\circ$ the \textsc{G-Tomo} extinction has a higher uncertainty because of steeper distance--extinction relations and spatial extinction variations. At high latitude $b>10^\circ$, or lines of sight where the extinction does not increase with distance, distance uncertainty has negligible effect. The \textsc{G-Tomo} uncertainties we present are based on distance uncertainty only. Note that \textsc{G-Tomo} extinctions can only be $\ge 0$: this gives a bias at the lowest extinctions in the plot. 

Considering only the high latitude sources for which we expect \textsc{G-Tomo} to be more accurate, there remain a number of outliers where the \textsc{PySSED} extinctions are above or below the \textsc{G-Tomo} predictions. Checking these sources, we see no indications of a higher uncertainty in the \textsc{PySSED} fits. The two such objects where \textsc{PySSED} most significantly finds a smaller extinction are MPA J1906-1634 and PHR J1911-1546, which both lie towards the Galactic centre region at latitudes of 10 and 15 degrees. These could be affected by the same issues as the low latitude sources in the Galactic plane. The sources offset above the relation include Abell\,30, which is known to have a dusty nebular core \citep{1994ApJ...435..722B}, and Abell\,16, a known binary. For Abell\,16, we see some IR excess but no report of central dust; the PySSED fit is good and there is no obvious reason for the $\sim 0.12$\,mag offset in $E(B-V)$.

The extinctions from \textsc{PySSED}, \textsc{G-Tomo}, \citetalias{2016MNRAS.455.1459F} and \citet{1998ApJS..117..361C} are consistent with one another. The \textsc{G-Tomo} extinctions show the poorest correlation within this set, which likely reflects a dependence on PNe distance determinations, which may still have significant uncertainties; and the angular resolution limits to which dust maps can be determined.

\subsection{Extinction laws}

As discussed in Sec.~\ref{sec:ext_laws}, the major extinction laws of \citetalias{1989ApJ...345..245C} and \citetalias{1999PASP..111...63F} tangibly disagree at the wavelength of H$\alpha$, thus predict different H$\alpha$/H$\beta$ ratios for the the same $E(B-V)$. In principle, comparing observed Balmer ratios and PySSED extinctions can identify which extinction law is more accurate at H$\alpha$, provided the data is sufficiently accurate and that internal extinction is negligible.

\begin{figure}
\includegraphics[width=\columnwidth]{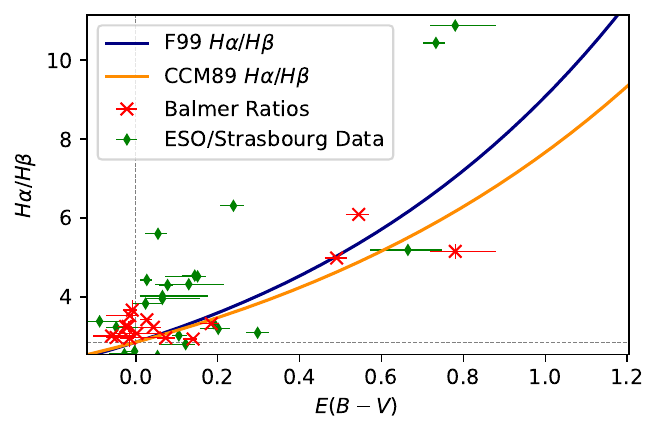}
\caption{Comparison of literature ${\rm H}\alpha/{\rm H}\beta$ ratios to our $E(B-V)$ results using \citetalias{1999PASP..111...63F}. Balmer ratios are taken directly from the original paper or calculated from the extinction and the extinction curve used in the paper, see Table \ref{tab:HaHb}. Data from the old ESO/Strasbourg catalogue is also presented \citep{1992secg.book.....A}. Slopes showing the predicted ${\rm H}\alpha/{\rm H}\beta$ for a given $E(B-V)$ are included for both the \citetalias{1999PASP..111...63F} and \citetalias{1989ApJ...345..245C} extinction laws.}
\label{fig:Ext_Law_Comp}
\end{figure} 

There are limited H$\alpha$/H$\beta$ data for our complete sample, probably because accurate spectra of faint, extended nebulae can be difficult to obtain.  We have compiled Balmer ratios for $32$ PNe in our sample, based on data published in or later than \citet{1994MNRAS.271..257K}; these are  listed in Table \ref{tab:HaHb}. We find that older data have uncertainties too large to be useful.

\begin{table*}
    \centering
    \caption{Observed H$\alpha$/H$\beta$ in the literature for PNe in the current sample. A H$\alpha$/H$\beta$ range is given when the literature data contains multiple observations of the nebula, an uncertainty is given when the literature quotes an uncertainty. The data included is all in or later than \citet{1994MNRAS.271..257K}.}
    \begin{tabular}{llccl}
    PN G & PN Name & H$\alpha$/H$\beta$ & $C_\beta$ &  Reference \\
    \hline
      093.4+05.4	&NGC 7008    &  4.5    & 0.55 & \cite{2010ApJ...711..619M} \\
       081.2-14.9	& Abell 78  &    2.87 - 2.99    & 0.09$\pm$0.03$^a$  & \cite{2005MNRAS.357..548P} \\
       204.0$-$08.5	& Abell 13  & 4.95 - 5.02 & $0.78 \pm 0.01$ & \cite{2005MNRAS.357..548P} \\
       302.1+00.3	& RCW 69  & 4.58 - 5.43   & 0.75$\pm$0.12 & \cite{2006MNRAS.372.1081F} \\
       035.9$-$01.1 &	Sh 2-71  & 4.97 - 5.34 &  0.82$\pm$0.05 & \cite{2001RMxAA..37..237B} \\
       208.9$-$07.8	& TaWe 1  & 3.32 & 0.2 & \cite{1999NewA....4...95A} \\ 
       222.1+03.9	& PFP 1  & 3.01 & 0.07 & \cite{2004PASA...21..334P}  \\
       217.1+14.7	& Abell 24   & 3.25 & 0.20  & \cite{2003RMxAA..39..149B} \\ 
       011.3$-$09.1 &	PTB 32   & 5.16  &   0.74     &  \cite{2006MNRAS.367.1551B}    \\
    165.5$-$15.2&	NGC 1514  & 5.37 & 0.88$\pm$0.07   & \cite{2021MNRAS.504.4806A} \\
       214.9+07.8	& Abell 20   & -- & 0.0  & \cite{2004AA...423..199C} 
       \\
       208.5+33.2	& Abell 30   &  -- &  0.0 & 
       \cite{1994MNRAS.271..257K} \\ 
       318.4+41.4	& Abell 36  &  3.23   & 0.17  &                   \cite{2015MNRAS.446..317A} \\
    047.0+42.4 &	Abell 39   & 2.96 $^b$ &   0.05$\pm$0.04 &       \cite{2001ApJ...560..272J} \\
        059.7$-$18.7	& Abell 72     & 3.42 &     0.25 & \cite{2005MNRAS.357..548P} \\  
        259.1+00.9	& Hen 2-11    & 16.81 &    2.41$\pm$0.01  &     \cite{2014AA...562A..89J} \\
    002.7$-$52.4	& IC 5148/50   & 3.52 &    0.38 &              \cite{1994MNRAS.271..257K}  \\ 
    164.8+31.1	& JnEr 1 & 3.40, $3.76\pm0.17$  &   0.26 , $0.34 \pm 0.31$ & \cite{2001RMxAA..37..237B}; \cite{2009ApJ...694.1335D} \\  
    294.1+43.6 &	NGC 4361    & 2.97  & 0.096$\pm$0.031 &
     \cite{2024AA...690A.264W} \\
     307.2$-$03.4	& NGC 5189   & 4.63 &   0.65 & \cite{1994MNRAS.271..257K} \\ 
     060.8$-$03.6	& NGC 6853   & $2.7\pm0.4$ & 0.0 & \cite{2010ApJ...711..619M} \\
     036.1$-$57.1	&NGC 7293    & 2.72 - 3.18  &   0.04$\pm$0.04   & \cite{1999ApJ...517..782H} \\ 
     258.5$-$01.3	& RCW 24     & 3.15 - 3.31  &    0.16$\pm$0.04   & \cite{2006MNRAS.372.1081F} \\   
     148.4+57.0	& NGC 3587 & $3.25\pm 0.14$ &    0.16$\pm$0.26  & \cite{2009ApJ...694.1335D} \\
     047.1$-$04.2	& Abell 62  & 3.31  &   0.20   &                \cite{2005MNRAS.357..548P} \\
     342.0$-$01.7	& PHR J1702$-$4443 &  11.2  &   1.81 & \cite{2015AA...583A..83A} \\ 
     247.8+04.9	  & FP J0821$-$2755  & 6.87 - 6.89  & 1.02$\pm$0.03   & \cite{2000AA...356..274Z} \\
     231.8+04.1	& NGC 2438  & 4.67, 3.3  &  0.67, 0.23$\pm$0.02 $^c$ &      \cite{1994MNRAS.271..257K,2014AA...565A..87O} \\   
     345.3$-$10.2	& MeWe 1-11   & 2.22 - 2.69  &     0.0   &  \cite{2004AA...423.1017E} \\ 
     216.0$-$00.2	& Abell 18  & -- &   1.41 & \cite{2004AA...423..199C} \\   
  \hline
    \end{tabular}
    \label{tab:HaHb}

    $^a$Abell 78 is found to have an electron temperature $\approx 18,000$ so an intrinsic $H\alpha / H\beta$ ratio of 2.75 is used. $^b$This ratio includes the \textsc{He II} line. $^c$The lower extinction is measured towards the halo of the PN.
\end{table*}

The comparison of the two laws is shown in Fig.~\ref{fig:Ext_Law_Comp}. Two data sets are shown: the observed Balmer ratios of \citet{1992secg.book.....A} and the compilation of recent measurements in Table \ref{tab:HaHb}. The $E(B-V)$ determinations in the figure are for  A/A+ class CSPNe candidate whose SEDs have at least five data points. This leaves rather few sources. High-extinction sources often have fewer data points, and some sources are absent as they are binaries for which we do not determine extinction. 

The two model extinction curves of Fig.~\ref{fig:Ext_Law_Comp} separate at higher extinctions. The older extinction values of the ESO/Strasbourg catalogue \citep{1992secg.book.....A} are clearly too inaccurate to distinguish the curves, whilst the compilation in Table \ref{tab:HaHb} follows the curves more closely but lacks sufficient PNe at high extinctions. 
There is currently insufficient high quality data to distinguish the curves.
New observations of the Balmer line ratios of high extinction PNe would be desirable.

\subsection{Circumnebular extinction}
\label{sec:internal}

Planetary nebulae are strong dust emitters, from dust produced within the stellar winds. This dust contributes significant extinction both on the AGB and during the post-AGB phase but the extinction diminishes as the nebula expands.

Circumnebular extinction is an important factor in the planetary nebula luminosity function (PNLF). The PNLF is based on the [O\,III] luminosities of PNe in external galaxies \citep{2025ApJ...983..129J}. These show a well-defined cut-off at high luminosities, which is sufficiently invariant to have become an important distance indicator for both spiral and elliptical galaxies. Evolutionary models have difficulty explaining this invariance \citep{2018NatAs...2..580G}: the most plausible model is that the most luminous PNe also have the highest internal extinctions, which compensates for the higher photon flux \citep{2025ApJ...983..129J}.

Our sample of more extended PNe is not expected to include these brightest nebulae, since the majority are already on the cooling track with greatly reduced stellar luminosities, but it may include objects that have evolved from such nebulae. Using the equation to calculate circumstellar PNe extinction presented in the supplementary discussion section of \citet{2018NatAs...2..580G}, for the smallest nebulae in our sample, which have diameters of order 0.25\,pc, we expect an expected circumnebular extinction $A_V\sim 0.13$\,dex corresponding to $E(B-V)\sim0.04$. This is near the limit of what we are able to confidently detect from comparison of our measured $E(B-V)$ results with those of \textsc{G-Tomo}. 

The extinction results of NGC 6781 are notable. With a physical minor axis of 0.26\,pc, it is one of the smallest nebula in the sample. If we assume that the difference between the \textsc{G-Tomo} extinction of $C_\beta=0.45 ^{+0.01}_{-0.02}$ and the \textsc{PySSED} extinction of $C_\beta=0.79^{+0.03}_{-0.05}$ is driven by circumnebular extinction, we may use the $R^{-1}$ relation of \citet{2018NatAs...2..580G} to find the expected circumnebular extinction when the nebula was at the size of the smallest Galactic PNe, which have diameters of order 0.06\,pc. This results in a prediction for the internal $C_\beta=1.04$, ($E(B-V)=0.65$) which is a factor of two above the extinctions that \citet{2025ApJ...983..129J} require to explain the cut-off in the PNLF. The extinction is not uniform across this source, and if we assume only half the source is extincted (e.g., from an inclined torus) then the resulting effective $C_\beta=0.26$, consistent with the PNLF requirements. This comparison suggests that the 3D structure of the circumstellar extinction may be important for the correct evaluation of its effect on the PNLF.

\section{Conclusions}

The goal of the project is to use a wide range of photometry of CSPNe to determine extinctions towards PNe. All 262 confirmed PNe in HASH with  minor axis diameter larger than 1 arcmin are investigated. Extinctions are derived by fitting TMAP stellar atmosphere models with $T_{\rm efff}=100\,$kK to the available photometry. For stars with $T_{\rm eff} > 60$\,kK, the slope of CSPNe spectra are found to be weakly dependent on temperature or other parameters. At lower temperatures, models at appropriate temperature are needed. 
For the sources with most accurate and comprehensive broadband photometry, $E(B-V)$ can be determined to $\pm 0.02$ at low extinctions.  In this complete sample of extended nebulae, we derive $E(B-V)$ values for 162 PNe which have good CSPNe candidates.

The extinction measurements of this paper 
are specific, independent of distance determination, and can probe extinction beyond the bounds of 3D galactic dust maps. The measurements are relatively homogenous, with some heterogeneity introduced by the variation and quality of available photometry. 

The results correlate well with \textsc{G-Tomo}, a 3D map of interstellar extinction, and with the compilation by \citetalias{2016MNRAS.455.1459F}. The agreement with older observations of PN extinctions is weaker, and is likely limited by the accuracy of those data.
The agreement between our results and \textsc{G-Tomo} indicates that extinction to our sample of extended PNe is dominated by interstellar dust. This indicates that any excess extinction dissipates as PNe evolve. Nonetheless, we find evidence of internal extinction for NGC 6781, consistent with the levels of circumnebular extinction required by \citet{2025ApJ...983..129J} to explain the bright end cut-off of the PNLF if the dust layer morphology is taken into account.

Luminosities were calculated based on an assumed stellar temperature of $T_{\rm eff} = 100$\,kK, which introduces an uncertainty of $\pm 0.5$\,dex in luminosity relative to the true temperature. For sources where a temperature is determined in the literature, we correct these luminosities to that temperature. These have much more precisely determined luminosities, but may be limited by distance and temperature uncertainty on a source by source basis. The majority of stars fall on the post-AGB cooling tracks of \citet{2016A&A...588A..25M}. 

The CSPNe of Abell 46 and HaWe 13 have been classified as post-RGB objects in the past. However, the uncertainties in the stellar temperatures are such that a location consistent with the post-AGB tracks cannot be excluded. For a third proposed post-RGB star, HaTr 4, we find that the star fits the post-AGB tracks.  The post-RGB classification for these three stars cannot be confirmed.

The data we collect enable us to thoroughly evaluate the CSPNe candidates of our sample. In many cases, we agree with the literature determinations, but we reject a few CSPNe candidates where we find their photometry incompatible with being a CSPN. We also identify eight new potential CSPNe candidates, mostly of newly discovered PNe, including three candidates with high confidence. If any CSPN misidentifications remain in our sample, they will introduce errors and limit the accuracy of our extinction results.

The extinction laws of \citetalias{1989ApJ...345..245C} and \citetalias{1999PASP..111...63F}  differ notably at the wavelength of H$\alpha$. In principle, comparing the stellar extinctions with accurate H$\alpha$/H$\beta$ ratios can distinguish these two laws. For these extended PNe, we find that  the available literature ratios are too limited and  may not be accurate enough. Further observations will be required in order to test the extinction laws.

Our sample contains seven stars with composite SEDs.  Four of these appear to be new binary identifications: We 3-1, PTB 32, Pa 161 and StDr 141. In the case of one PN with a previously identified binary, K\,1-6, we find indications that either the identified binary pair is foreground to the true CSPN, or the nebula is not a true PN. 

The binary fraction in our full sample of 162 A-class stars is 23\%. This increases to 36\% for objects with Galactic latitudes $b>10^\circ$. The differences indicate a detection bias. Some companions stars remain difficult to detect, especially low-mass main sequence stars and white dwarfs on wide orbits. The fraction is not inconsistent with the binary fractions of main-sequence stars.

\section*{Acknowledgements}

This research acknowledges financial support from the European Union through the Open Science Clusters' Actions for Research \&\ Society (OSCARS) project 01-358, funded through HORIZON-INFRA-2023-EOSC-01-01.  STFC/UKRI support was obtained through a PhD studentship and through through grar=nts ST/T000414/1 and  ST/X001229/1. A.A.Z. aacknowledges support from the Royal Society through grant IES/R3/233287 and 
from the University of Macquarie.

The TheoSSA service (\url{http://dc.g-vo.org/theossa}) used to retrieve theoretical spectra for this paper was constructed as part of the activities of the German Astrophysical Virtual Observatory.

This research has made use of the HASH PN database at \url{hashpn.space}.

\section*{Data Availability}

All data used in this project is publicly available through Vizier and TheoSSA. The base PySSED software is available from  \url{https://github.com/iain-mcdonald/PySSED} and can also be accessed in single-source mode at \url{https://explore-platform.eu} under the name 'S-Phot'.



\bibliographystyle{mnras}
\bibliography{biblio} 




\appendix

\section{Binary SEDs}
\label{app:binary_SEDs}

Figures A1--A4 show binary fits to the central systems of four PNe which are newly identified as binaries in this paper. Extinction cannot be measured to these binary systems, so the \textsc{G-Tomo} extinctions are adopted instead. In the top axis of each figure, a hot CSPN is reddened by the \textsc{G-Tomo} $E(B-V)$ then fitted to the UV data points. The predicted emission of the CSPN at the optical data points is then subtracted from the observed fluxes through these filters. The subtracted fluxes are then dereddened and fit using the simple main sequence star fitting procedure of \textsc{PySSED}. The resulting fit is shown in the lower axis. Observations which have been masked and left unused by the simple fit procedure are shown in grey. 

\begin{figure}
{\includegraphics[width=\columnwidth]{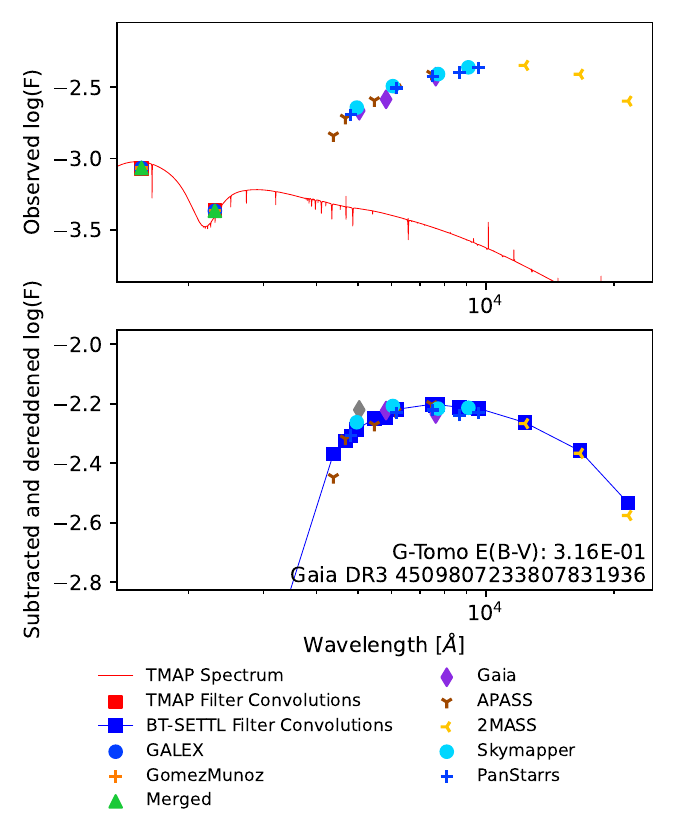}}
\caption{We 1-3 Binary Fit} 
\label{fig:We1-3_fit}
\end{figure}

\begin{figure}
{\includegraphics[width=\columnwidth]{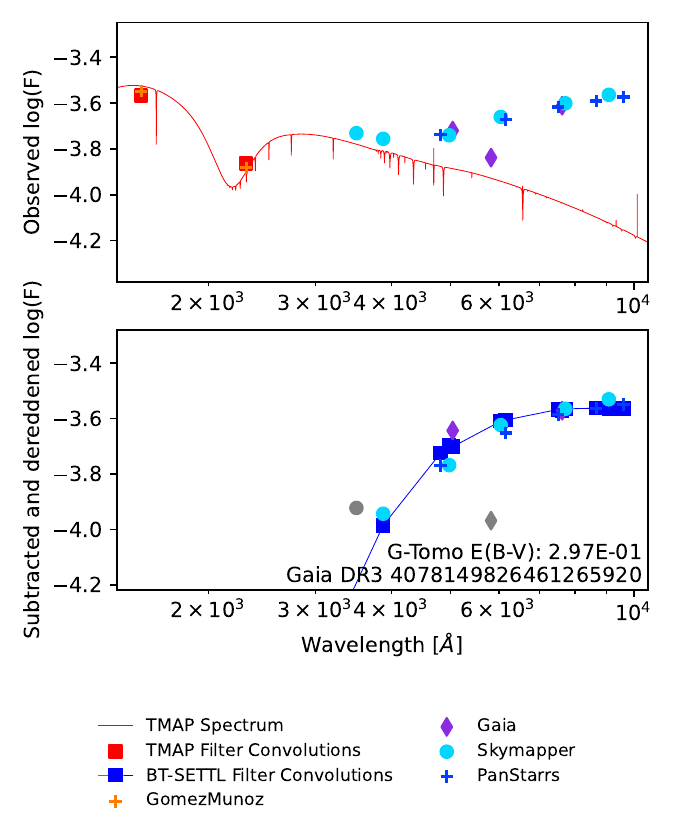}}
\caption{PTB 32 Binary Fit} 
\label{fig:PTB32_fit}
\end{figure} 

\begin{figure}
{\includegraphics[width=\columnwidth]{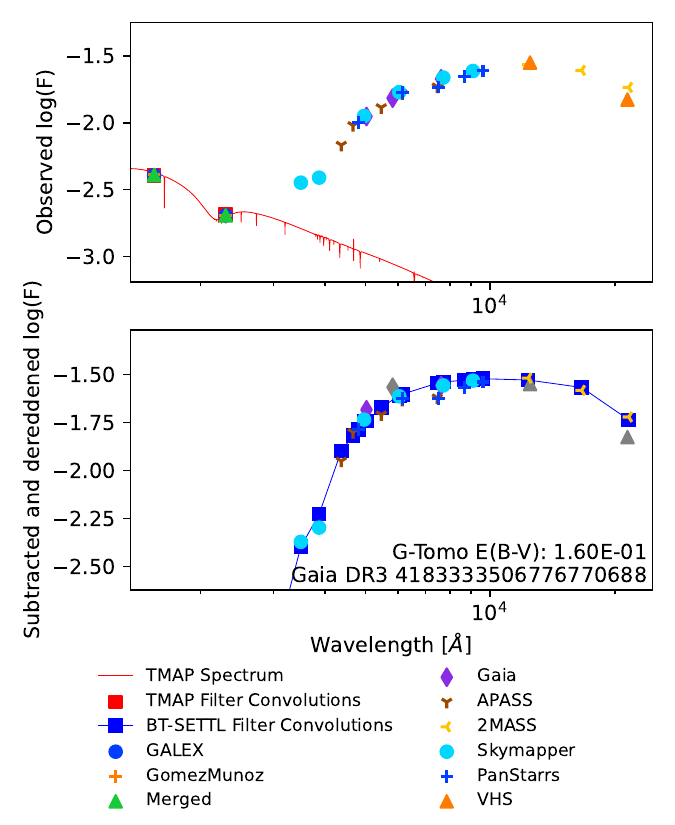}}
\caption{Pa 161 Binary Fit} 
\label{fig:Pa161_fit}
\end{figure}

\begin{figure}
{\includegraphics[width=\columnwidth]{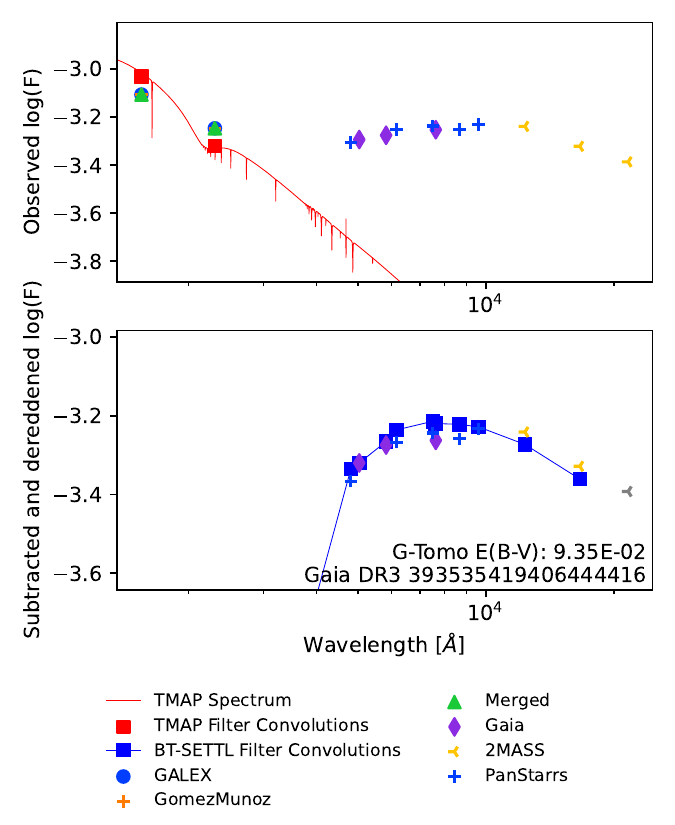}}
\caption{StDr 141 Binary Fit} 
\label{fig:StDr141_fit}
\end{figure} 

\section{Excluded systems}
There are two exceptional cases within our sample of 262 objects which needed to be excluded from our analysis by giving their CSPN candidates the exceptional F* grade. For LoTr 5, a known binary system, the observed photometry of the supergiant companion to the CSPN dominates the photometry wavelengths into the UV, making it unsuitable for our method. For BMP J1808-1406, we find reason to doubt its PN status. The \textit{Gaia} parallax is uncertain, and using the distance from \citetalias{2016MNRAS.455.1459F} gives the identified CSPN an extreme low luminosity of $10\,\rm L_\odot$. Given a general lack of data on the object, we are not certain that the PN status is robust and recommend further observations.


\bsp	
\label{lastpage}
\end{document}